\newif\ifSUBMIT%
\newif\ifCOMMENTS%
\let\ifCOMMENTS\iffalse{}
\let\ifSUBMIT\iffalse{}
\let\ifSUBMIT\iftrue{}

  \newlength{\figurewide}
\documentclass[%
  preprint,11pt,   % Single Column: use ToEditSingleColumn.tex too
  aps,prb,amsmath,amssymb,amsfonts,
  showkeys,superscriptaddress,%
  floatfix,
]{revtex4-1} % chktex 8
  \usepackage{graphicx}

\ifSUBMIT%
  \ifCOMMENTS%
    \usepackage{color}    %%% NOTE this is necessary in pdflatex
    \usepackage[normalem]{ulem}

    \def\NSTRIKE#1{{\color{blue}\sout{#1}}}
  \else

    \def\NSTRIKE#1{}
  \fi
\else
  \usepackage{color}    %%% NOTE this is necessary in pdflatex
  \usepackage[normalem]{ulem}
  \def\NSTRIKE#1{{\color{blue}\sout{#1}}}
\fi

  \AtBeginDocument{\usepackage{booktabs}}
    \usepackage{physics} %common notations for physics
    \usepackage{bm} %bold math
    \usepackage[version=4]{mhchem} % chemical equation
    \newcommand{\kB}{\ensuremath{k_\mathrm{B}}}
    \newcommand{\ii}{\ensuremath{\mathrm{i}}}
    \newcommand{\ee}{\ensuremath{\mathrm{e}}}
    
  \usepackage{acronym} % Provides for acronyms
    \acrodef{DoF}{degrees of freedom}
    \acrodef{EoM}{equations of motion}
    \acrodef{IRC}{intrinsic reaction coordinate}
    \acrodef{MEP}{minimum energy path}
    \acrodef{NHIM}{normally hyperbolic invariant manifold}
    \acrodef{TS}{transition state}
    \acrodef{TST}{transition state theory}
    \acrodef{PODS}{periodic orbit dividing surface}
    \acrodef{CPT}{canonical perturbation theory}
    \acrodef{LCS}{Lagrangian coherent structure}
    \acrodef{LD}{Lagrangian descriptor}
    \acrodef{RDS}{random dynamical system}
    \acrodef{SDE}{stochastic differential equation}
    \acrodef{RDE}{random differential equation}
    \acrodef{ODE}{ordinary differential equation}
    \acrodef{ATI}{asymptotic trajectory indicator}
    \acrodef{NBC-ATI}{neighbor bisection and continuation with ATI}

\begin{document}

\title{Identifying Reaction Pathways in Phase Space via Asymptotic Trajectories}
  \author{Yutaka Nagahata}
  %  \email{ynagaha1@jhu.edu}
    \affiliation{%
      Department of Chemistry,
      Johns Hopkins University,
      Baltimore, MD 21218
    }
  %\author{Florentino Borondo}
  \author{F. Borondo}
    \affiliation{%
      Instituto de Ciencias Matem\'{a}ticas (ICMAT),
      Cantoblanco, 28049 Madrid, Spain
    }
    \affiliation{%
      Departamento de Qu\'{\i}mica,
      Universidad Aut\'{o}noma de Madrid,
      Cantoblanco, 28049 Madrid, Spain
    }
  %\author{Rosa Mar{\'{\i}}a Benito}
  \author{R. M. Benito}
    \affiliation{%
      Grupo de Sistemas Complejos,
      Escuela T\'{e}cnica Superior de Ingenier\'{\i}a Agron\'{o}mica,
      Alimentaria y de Biosistemas,
      Universidad Polit\'{e}cnica de Madrid, 28040 Madrid, Spain
    }
  \author{Rigoberto Hernandez}
    \email{r.hernandez@jhu.edu}
    \affiliation{%
      Department of Chemistry,
      Johns Hopkins University,
      Baltimore, MD 21218
    }
  \date{\today}
\preprint{\it Phys. Chem. Chem. Phys.\/}

\begin{abstract}
  In this paper, we revisit
  the concepts of the reactivity map and the reactivity bands
  as an alternative to the use of perturbation theory for the
  determination of the phase space geometry of chemical reactions.
  We introduce a reformulated metric, called the asymptotic trajectory indicator,
  and an efficient algorithm to obtain reactivity boundaries.
  We demonstrate that this method has
  sufficient accuracy to reproduce phase space structures
  such as turnstiles for a 1D model of the isomerization of ketene in an
  external field.
  The asymptotic trajectory indicator can be applied to
  higher dimensional systems coupled to Langevin baths as we demonstrate
  for a 3D model of the isomerization of ketene.
\end{abstract}

\maketitle

\section{Introduction\label{sec:Intro}}

  The identification of a reaction path (or pathway)
  has received attention
  from the beginning of the development of \ac{TST}\cite{eyring35,Evans1935,wigner38}
  to characterize the energetics of the reaction between reactants and products.
  Eyring called it 
  ``the path requiring least energy''\cite{Glasstone1941} 
  which is now commonly called the \ac{MEP}
  and obtained as the path of least resistance starting from the energy minimum associated
  with the reactant.
  Fukui is generally credited for what came to be known as the \ac{IRC}\cite{Fukui1970,Fukui1976}
  because he introduced it, in a mass-weighted coordinate system, 
  as the path of steepest descent starting from the
  saddle.
  Beyond the developments in the statistical formulation of
  \ac{TST},\cite{Glasstone1941,Laidler1983,truh83,rmp90,truh96}
  the re-imagination of the reaction path
  as an object in full phase space\cite{keck67}
  led to the use of reactivity bands\cite{Wall1958,Wright1978}
  and the \ac{PODS}\cite{pollak78}
  to characterize reactions.
  Both analyses are formulated in terms of
  {identifying} the trajectories in between the reactive and non-reactive
  trajectories. 
  See Ref.~\onlinecite{komatsuzaki13a} for more details, and
  Ref.~\onlinecite{keshavamurthy18}
  for the connection to reaction path sampling\cite{chan98a,chan02b}
  also qualitatively described in our earlier work.\cite{dawn05a,hern10a}
  We are thus led to use these mathematical structures to reconsider 
  the determination of the optimal reaction path {in phase space}.

  In the 1980s, dynamical system theory was advanced through the use of
  the Poincar{\'{e}} map,\cite{Davis1984}
  turnstile\cite{Mackay1984} structures,
  and reaction island theory.\cite{deLeon1}
  Unfortunately, all 
  of these methods are applicable mostly
  to systems up to 2 \ac{DoF}.
  Going beyond this restriction,
  Wiggins\cite{Wiggins1990} suggested a multidimensional
  generalization of the unstable periodic orbit in 2 \ac{DoF} systems
  and the reaction pathway associated with it.
  The former is \iac{NHIM},
  and the latter is the boundary of the reaction pathway 
  which can be understood
  as the stable and unstable manifolds of the NHIM\@.
  In the limit of two dimensional systems, the \ac{NHIM} 
  is an unstable periodic orbit
  which was later understood as an anchor of 
  the \ac{PODS}.\cite{wiggins01,hern10a}
  Fenichel\cite{Fenichel72}
  was the first to prove that these \acp{NHIM}
  persist under perturbation.
  For simplicity, in this work, we call this and its 
  subsequent generalizations, the NHIM Persistence Theorem
  (see Appendix~\ref{apx:NHIMtheorem}).
  A consequence of this theorem is that
  the saddle point on the potential energy surface plays
  a significant role in many cases because
  the structure of the \ac{NHIM} near the energy of the saddle point will persist 
  for larger energies
  as long as the truncated higher-order terms in perturbation theory are small enough.

  The pioneering work on the application of perturbation theory
  for reaction dynamics in the chemical physics community was made
  by Hernandez and Miller\cite{hern93b,hern94}
  through Van Vleck semiclassical perturbation theory
  and by Komatsuzaki et al.\cite{Komatsuzaki1996}
  through classical Lie-\ac{CPT}
  (also known as {a component of} Birkhoff normal form theory)
  to obtain non-recrossing dividing surfaces in many \acp{DoF}.
  Uzer et al.~\cite{Uzer02} {later} showed the relation between 
  {this} normal form theory
  and the geometry of {the} reaction pathway.
  The normal form theory was also further generalized to address
  related challenges in
  quantum systems,\cite{Waalkens2008}
  the effects under rotational coupling,\cite{Kawai2011_,Ciftci2012}
  Langevin dynamics,\cite{Kawai2009a,Kawai2009b}
  generalized Langevin dynamics,\cite{Kawai2010c}
  and the classical\cite{Kawai07}
  and the quantum\cite{Kawai11laser}
  dynamics under an external field.
  The extraction of the \ac{TS} trajectory\cite{dawn05a}
  and relaxation of the normal form theory\cite{komatsuzaki06a,Komatsuzaki2010}
  plays a crucial role especially in time-dependent\cite{Kawai07,Kawai11laser}
  and stochastic\cite{hern06d,Kawai2009a,Kawai2009b,Kawai2010c} theories.
  Naturally, perturbation theories are limited and are applicable only when
  the zeroth order approximation (typically normal mode Hamiltonian) is valid,
  and the asymptotic series returns converged results.

  Challenges to this theory arise when
  the \ac{TS} or the phase space bottleneck
  is not strongly dominated by a potential energy saddle point.
  Perturbation theories expanded around such a point
  are not expected to be accurate unless
  the bottleneck happens to remain within
  the convergence radius of the perturbation.\cite{komatsuzaki06a}
  One possible such challenge comes from the existing of
  the roaming reaction pathway observed experimentally
  in formaldehyde \ce{H2CO} decomposition path to \ce{H_2 + CO}.\cite{bowman04a,bowman2011}
  The phase space manifestation of this system\cite{Mauguiere2017}
  shows the significant role of the reactivity boundary
  in the absence of a potential energy saddle.
  Nevertheless, the structure of transition state theory can be
  preserved even when such roaming reactions are present through
  the identification and use of a global transition state dividing surfaces.\cite{hern13c}
  There are other known examples that the
  reaction mechanisms are irregular due to factors such as
  long-time trapping in a well,\cite{Davis1984}
  bifurcation of the \ac{PODS},\cite{pollak78,komatsuzaki06a}
  dynamical switching of the reaction coordinate,\cite{komatsuzaki11}
  and the presence of 
  near higher-index saddles,\cite{Nagahata2013a,komatsuzaki13a}
  {with chemical species shown in the references.}
  In the limit of two or fewer \acp{DoF}, 
  there are nonperturbative approaches
  such as periodic orbit analysis,
  which is used in an effectively 2 \ac{DoF} system of the roaming reaction path:
  \ce{H2CO -> OC\bond{~}H_2} {\ce{-> H_2 + CO}}.\cite{Mauguiere2017}
  However,
  the extension of these approaches to the systems with higher than 2 \ac{DoF}
  is still a challenging task.

  An alternative approach to determining
  {the reactivity boundary}
  is rooted in
  the \ac{LCS}\cite{Haller2015,haller17a}
  of the dynamics in the Lagrangian frame.
  The \ac{LCS} is mediated by
  the \ac{NHIM} and its stable/unstable manifolds, and
  can be evaluated by numerical analysis\cite{haller17a}
  ---{\it e.g.}, through the identification of
  the finite time Lyapunov exponent (FTLE) ridge.
  The FTLE analysis has mostly been employed
  in effectively two-dimensional systems,
  such as ocean flows
  due to theoretical and numerical limitations.
  The \ac{LD}\cite{Mancho2010}
  was proposed as a heuristic alternative.
  In the context of reaction dynamics,
  the \ac{LD} was first introduced by Craven and Hernandez.\cite{hern15e}
  The theory provides, for example, a way to obtain nonrecrossing dividing surfaces
  in barrierless reactions in which perturbation theory is nonsensical because
  no unique zeroth-order dividing surface is available.\cite{hern16a}
  It has also been used to reveal geometric features in molecular systems
  such as ketene\cite{hern16d} and LiCN.\cite{Revuelta19}

  \begin{figure}[ht!]
    \includegraphics[clip=true,width =\figurewide]{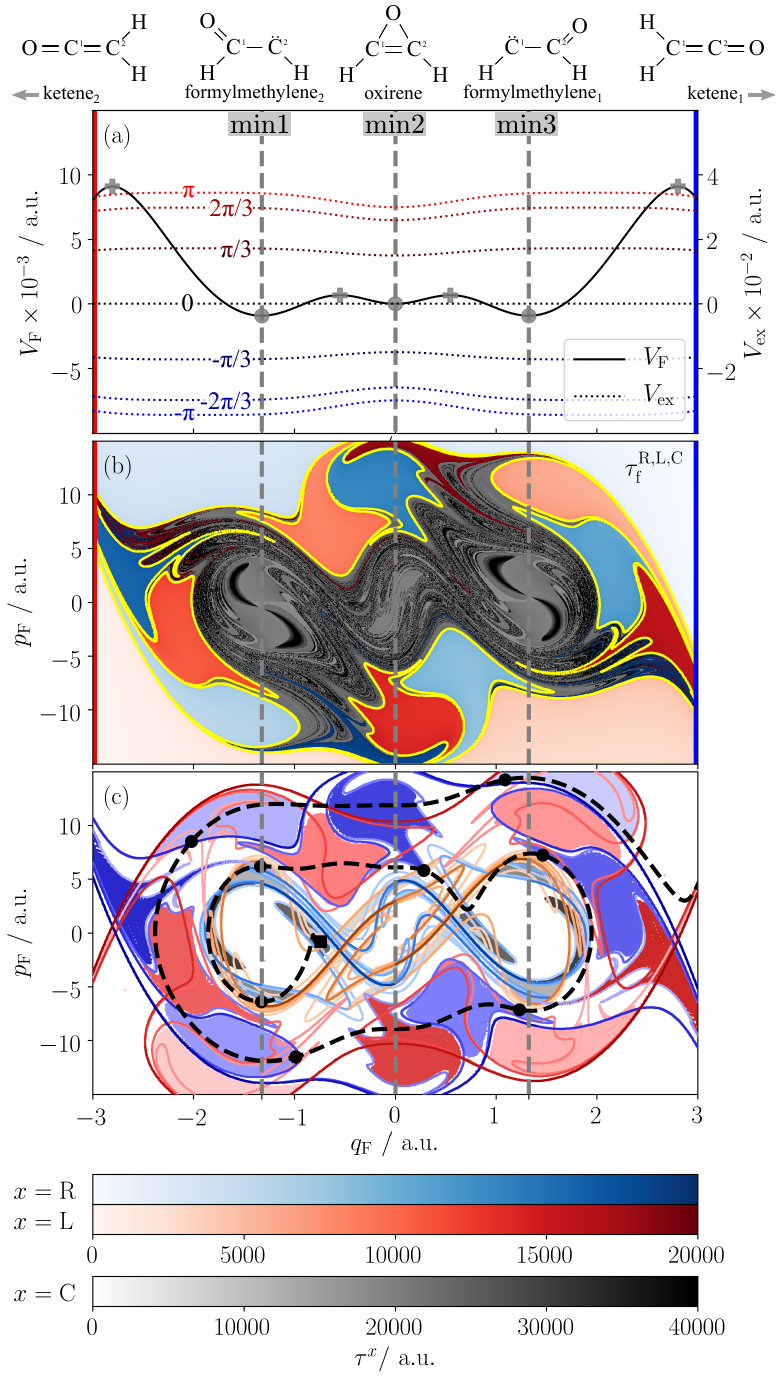}
    \caption{%
      Highlights of the main results of this paper for the
      1 D ketene
      isomerization reaction
      in an external field:
      (a)
      The potential energy $V_\mathrm{F}(q_\mathrm{F})$ (solid black line),
      the external force through dipole interaction
      $V_\mathrm{ex}(q_\mathrm{F}, t)$
      at phases $\omega t= n\pi/3$ for $n=3,2,\dots,-3$
      (colored dotted lines with values of $V_\mathrm{ex}$ shown in the right axis),
      the \acp*{TS} (gray plus symbols),
      the minima (gray filled-circles and vertical dashed lines) corresponding to
      the chemical formulas (ketene, formylmethylene, and oxirene),
      and the left (red) and right (blue) absorbing boundaries.
      (b)
      The first passage time to the left (right) absorbing boundary
      $\tau_\mathrm{f}^\mathrm{L}$ ($\tau_\mathrm{f}^\mathrm{R}$),
      or recurrent time after $\omega t = 14 \pi$ (gray),
      or \acs*{ATI} with the color scale given at the bottom.
      They provide the phase space skeleton of the reaction (yellow).
      (c)
      The phase space geometry structures shown here for comparison are
      the stable (blue) and unstable (red) manifolds,
      and regions whose points exit right (blue) or left (red).
      A sample trajectory (black dashed line) is shown to illustrate
      the points on the Poincar{\'e} surface of section in corresponding regions:
      square and circles on the time slice $|\omega t|\equiv 0 \pmod{2 \pi}$.
    }\label{fig:highlight}
  \end{figure}

  In this paper,
  we propose a non-heuristic approach for locating
  the reaction pathways in a phase space.
  The approach is an alternative to the nonperturbative methods cited above.
  Toward its formulation,
  we revisit the theory of the reactivity map\cite{Wright1978}
  and reactivity boundaries.\cite{Nagahata2013a}
  We introduce the \ac{ATI} associated with
  the reactivity map, and
  formulate it in the context of
  dynamical systems theory in Sec.~\ref{sec:Theory}.
  It is employed in a numerically efficient algorithm
  to extract the reactivity boundary in Sec.~\ref{sec:Method}.
  The analysis is applicable
  whenever the solutions of the equations of motion
  are continuous in some sense
  with respect to initial conditions and integration time.
  In the case of smooth Hamiltonians, this is automatically
  satisfied with respect to the standard definition of continuity,
  but in more general cases, such as with stochastic equations of motion,
  it is sufficient to define continuity with respect to neighborhoods
  of the input and output variables.
  Consequently, one can use \ac{ATI} even for systems coupled to a Langevin bath
  as discussed briefly in Subsec.~\ref{ssec:RDS}.

  In what follows, we demonstrate the \ac{NBC-ATI} method
  through application to the 1 \ac{DoF} and 3 \acp{DoF} reduced ketene
  models\cite{gezelter1995} 
  under an external field and a Langevin bath, respectively,
  in Sec.~\ref{sec:Application}.
  In the former case illustrated in Fig.~\ref{fig:highlight}a, 
  this method uses the \ac{ATI} shown in Fig.~\ref{fig:highlight}b to uncover
  the complex reaction pathway ---with respect to the reactivity boundary---
  and how it is guided by the phase space structure
  associated with the four potential energy saddles shown in
  Fig.~\ref{fig:highlight}a.
  As a result of this analysis, we find the rare reactive pathway
  between ketene and the intermediate structures
  ---that is, formylmethylenes and oxirene---
  shown as the  black dashed line in Fig.~\ref{fig:highlight}c.
  In general, the \ac{ATI} can be used to locate
  the phase space skeleton of the reaction,
  the stable and unstable manifolds of all available \acp{NHIM}.

\section{Theory}\label{sec:Theory}

\subsection{Phase Space Flow around an Index-one Saddle}
  \begin{figure}[t]
    \includegraphics[clip=true,width =\figurewide]{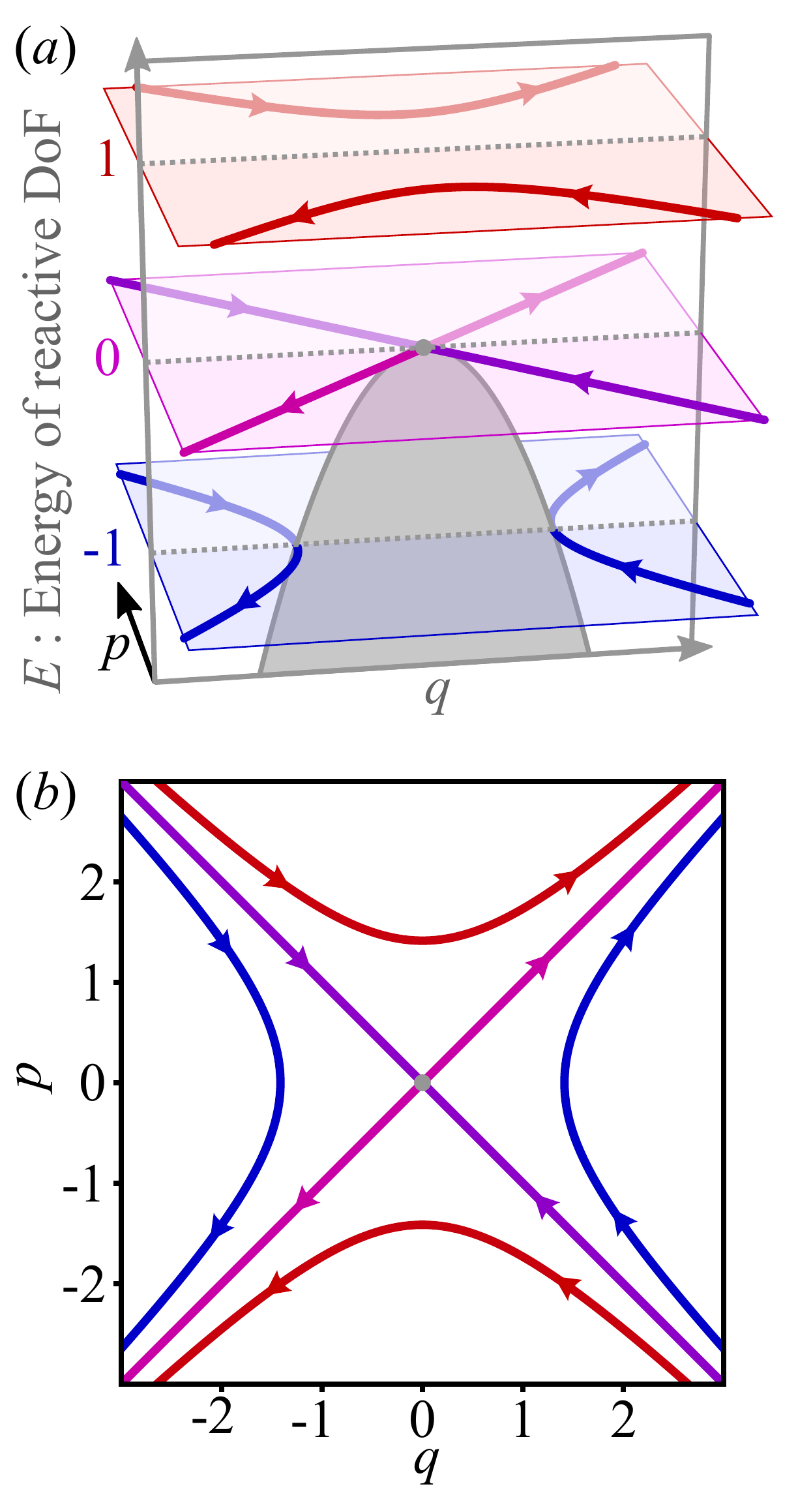}
    \caption{%
      The phase space flow of trajectories along a reactive \ac{DoF}
      is shown for various energies and its projection in phase space.
      For simplicity, the Hamiltonian, ${\cal H}=(p^2-q^2)/2$,
      of this hyperbolic normal mode is taken to be that 
      at second order in the potential expansion.
      Reactive trajectories (red, $E=1$),
      non-reactive trajectories (blue, $E=-1$), and
      trajectories on the reactivity boundary (purple and pink, $E=0$)
      in (a) the $q$-$p$-$E$ space, and (b) $q$-$p$ space.
      The potential energy surface is also drawn
      in gray on (a) $q$-$E$ plane.
    }\label{fig:HyperbolicFlow}
  \end{figure}
  For the normal mode approximation of a Hamiltonian expanded at an index-one saddle point:
  \begin{flalign}
    H_0(\bm{p},\bm{q})&=
    \frac{1}{2}(p_1^2-\omega_1^2 q_1^2)+
    \frac{1}{2}\sum_{i=2}^n(p_i^2+\omega_i^2 q_i^2), \label{eq:H0}
  \end{flalign}
  here we use $i$th normal mode coordinate $q_i$,
  its conjugate momentum $p_i$,
  its frequency $\omega_i$,
  the coordinates vector $\bm{q}=(q_1,\dots,q_n)$,
  and
  the momenta vector $\bm{p}=(p_1,\dots,p_n)$.
  For all $i$, $\omega_i$ is positive
  so that $\omega_1 \ii$ is a pure imaginary frequency.
  The index of a potential energy saddle point is
  determined by the number of Hessian eigenvalues,
  which correspond to the Morse index of a critical point.
  The Hamiltonian equation of motion can be written with $(\bm{p},\bm{q})$ by
  \begin{flalign}
    \dv{t} \mqty(p_i \\ q_i)
    = \mqty(\mp \omega_i^2 q_i \\ p_i),\label{eq:RCEoMNM}
  \end{flalign}
  where $\mp$ correspond to the sign of the monomial
  $\pm \omega_i^2 q_i^2$ in Eq.~\eqref{eq:H0}, respectively.
  The first normal mode $i=1$ is called hyperbolic or reactive \ac{DoF},
  and is shown in Fig.~\ref{fig:HyperbolicFlow} with $\omega_1=1$.
  In the upper panel (\ref{fig:HyperbolicFlow}a), 
  trajectories with given reactive mode-energies
  are shown relative to
  the corresponding potential energy saddle.
  When the mode-energy is positive and negative,
  the trajectory is reactive and non-reactive, respectively.
  The boundaries in between
  consist of asymptotic trajectories
  with zero mode-energy toward or from the origin.
  Here, the origin of the reactive \ac{DoF} is called the \ac{NHIM}, 
  and its dimensionality grows
  as the total number of \acp{DoF} is increased.
  The trajectories asymptotic toward or from the \ac{NHIM} are called
  the stable or unstable manifolds of the \ac{NHIM}, respectively.
  In Wigner's \ac{TST} formulation,\cite{wigner38}
  the dividing surface is defined as $q_1=0$ and $p_1>0$.
  Thus, the \ac{NHIM}: $(p_1,q_1)=(0,0)$ is known as an anchor of the dividing surface.
  For a normal mode Hamiltonian,
  one can solve the equation of motion by the finding constants of motion,
  i.e., normal mode action $\bm{J}$.
  One can rewrite the reference Hamiltonian as
  \begin{equation}
    H_0=\sum_{i=1}^n \omega_i J_i(p_i,q_i),
    \label{eq:H0action}
  \end{equation}
  where the $i$th generalized momentum is
  \begin{equation}
    J_i=
    \begin{cases}
      \frac{1}{2\omega_i}(p_i^2-\omega_i^2 q_i^2)~(i=1)\\
      \frac{1}{2\omega_i}(p_i^2+\omega_i^2 q_i^2)~(i\ge 2)\\
    \end{cases}.
    \label{eq:action}
  \end{equation}
  %In fact, this $J_i$ is $i$th generalized momentum,
  which is conjugate to the action $\theta_i$,
  and together satisfy
  \begin{flalign}
    \dv{t} \mqty(J_i \\ \theta_i)
    = \mqty(-\pdv*{H_0}{\theta_i} \\ \pdv*{H_0}{J_i})
    = \mqty(0 \\ \omega_i)
  \end{flalign}
  with its solution $\theta_i=\omega_i t, J_i=\mbox{constant}{.}$.

  If $H_0$ is dominant,
  a similar relation can be obtained from CPT\@.
  When one can expand the Hamiltonian at the index-one saddle point:
  \begin{equation}
    H(\bm{p},\bm{q})=H_0(\bm{p},\bm{q})
    +\sum_{k=1}\epsilon^k{V}_k(\bm{q}), \label{eq:Hsaddle}
  \end{equation}
  where
  $V_k(\bm{q})$ is a ($k+2$)-order polynomial in $\bm{q}$,
  and
  the perturbation order $k$ is tracked by $\epsilon=1$
  without changing the equation.

  The construction of perturbation theory now follows a series of canonical transformations
  that successively remove the terms in $\theta$ up to a desired order while
  formally preserving the Hamiltonian structure.
  That is, we seek to find the composite transformation from 
  $(\hat{\bm{p}},\hat{\bm{q}})$  to
  $(\hat{\bm{J}},\hat{\bm{\theta}})$ such that
  the new Hamiltonian is
  $H=\check{H}(\hat{\bm{J}})+\order{\epsilon^k}$.
  In Lie-\ac{CPT},
  this is achieved through a ``time'' propagation that go forward or backward in time
  resulting in the solutions, $\hat{F}$ and $\check{F}$, respectively.
  One can solve the equation of motion of $\check{H}$
  with the order of accuracy $\order{\epsilon^k}$, i.e.,
  \begin{flalign}
    \dv{t} \mqty(\hat{J}_i \\ \hat{\theta}_i)
    = \mqty(-\pdv*{\check{H}}{\hat{\theta}_i} \\ \pdv*{\check{H}}{\hat{J}_i})
    = \mqty(0 \\ \tilde{\omega}_i),
  \end{flalign}
  where $\tilde{\omega}_i:=\pdv*{\check{H}}{\hat{J}_i}$.
  This equation can be rewritten with $(\hat{\bm{p}},\hat{\bm{q}})$ as
  \begin{flalign}
    \dv{t} \mqty(\hat{p}_i \\ \hat{q}_i)
    = \frac{\tilde{\omega}_i(\hat{\bm{J}})}{\omega_i}
    \mqty(\mp \omega_i^2 \hat{q}_i \\ \hat{p}_i),\label{eq:RCEoMNF}
  \end{flalign}
  where the difference from Eq.~\eqref{eq:RCEoMNM} is the locally constant term,
  $\tilde{\omega}_i(\hat{\bm{J}})/\omega_i$.
  Therefore, the coordinate transformation
  $(\bm{p},\bm{q})\to(\hat{\bm{p}},\hat{\bm{q}})$
  gives a local (in the sense of $\order{\epsilon^k}$) independence
  for each \ac{DoF}
  including the reactive \ac{DoF} $(i=1)$.
  This suggests that the phase space flow of Eq.~\eqref{eq:RCEoMNF}
  has a similar shape with Fig.~\ref{fig:HyperbolicFlow},
  but in the space of $(\hat{\bm{p}},\hat{\bm{q}})$.
  Revisiting the flow of the trajectories without specifically constructing
  the Lie-CPT transformations should thus result in an alternate constrution
  revealing the reaction path, and serves to motivate the approach pursued here.

\subsection{Inhomogeneity of the \ac{LD} on the \ac{NHIM}}

  \begin{figure}[t]
    \includegraphics[clip=true,width =\figurewide]{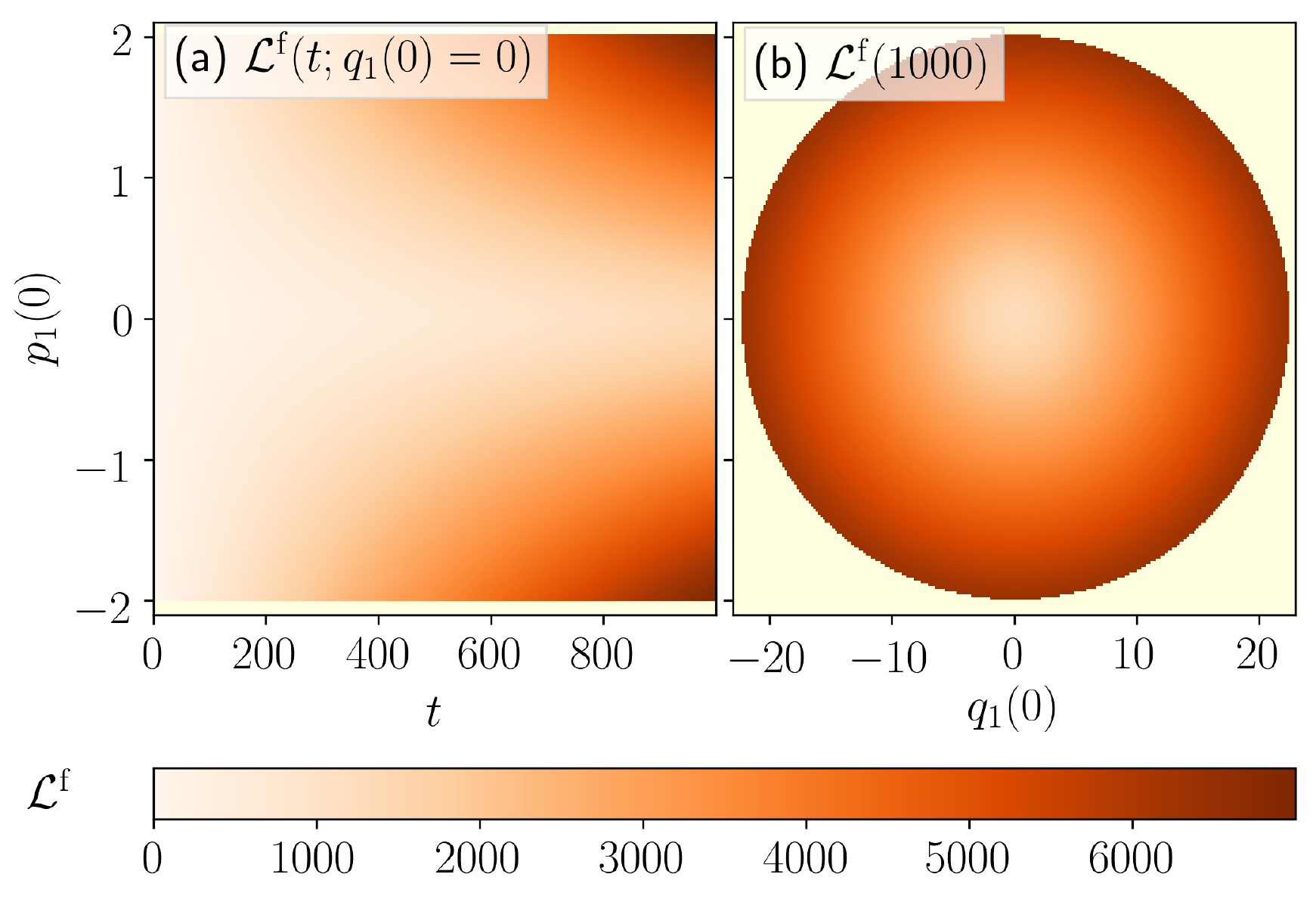}
    \caption{%
      Forward extremal Lagrangian Descriptor $\mathcal{L}^\mathrm{f}$
      (arc length of a trajectory) over initial conditions
      of a 2-\ac{DoF} harmonic system:
      $H=\sum_{i=1,2} p_i^2/(2m_i)+m_i k_i q_i^2/2$
      where $m_1=0.2,k_1=0.2,m_2=5,k_2=3$,
      with $E=10$, $q_2(0)=0$, and $p_2(0)>0$.
      The light-yellow area is energetically prohibited.
      (a) time evolution of $\mathcal{L}^\mathrm{f}(t)$ for $q_1(0)=0$
      in $t-p_1(0)$ space,
      (b) $\mathcal{L}^\mathrm{f}(t=1000)$ in $q_1(0)-p_1(0)$ space.
    }\label{fig:NM_LD}
  \end{figure}

  To obtain the \ac{NHIM} and its stable and unstable manifolds
  non-perturbatively,
  the \ac{LCS} and \ac{LD} are introduced here.
  In dynamical systems theory, a distinguished hyperbolic trajectory 
  has been defined\cite{wiggins02}
  as the non-autonomous analogue of a hyperbolic fixed point.
  Mancho and coworkers\cite{Mancho2009,Mancho2010}
  introduced the form of the \ac{LD} initially as a means 
  to locate these distinguished hyperbolic trajectories.

  Specifically in the context of the reaction dynamics,
  Craven and Hernandez\cite{hern15e} implemented
  the {\it extremal \ac{LD}}, $\mathcal{M}_\mathrm{ex}$, which
  describes the arc length of trajectories in coordinate space
  \begin{equation}
    \mathcal{M}_\mathrm{ex}(\bm{q}_0,\dot{\bm{q}_0},t_0;t):=
    \int_{t_0-t}^{t_0+t} \| \dot{\bm{q}} \| \dd t\;. \label{eq:LDdef}
  \end{equation}
  The arc length also evaluated for the forward and backward \acp{LD}:
  \begin{subequations}\label{eq:LDfLDb}
  \begin{eqnarray}
    \mathcal{L}_\mathrm{f}&:=&\int_{t_0}^{t_0+\tau} \| \dot{\bm{q}} \| \dd t
    \label{eq:LDf}
    \\
    \mathcal{L}_\mathrm{b}&:=&\int_{t_0-\tau}^{t_0} \| \dot{\bm{q}} \| \dd t
  \label{eq:LDb}
  \end{eqnarray}
  \end{subequations}
  to estimate stable and unstable manifolds, respectively,
  as the {\it ``abrupt change''\/}\cite{Mancho2013} of \acp{LD}.

  The \acp{LD}, $\mathcal{L}_\mathrm{f}$ and $\mathcal{L}_\mathrm{b}$,
  ---refer to Eq.~\ref{eq:LDfLDb}--- are
  accumulated value of a positive scalar along a trajectory.
  Trajectories that diverge from each other will necessarily accumulate
  different \acp{LD}, and the separation of these values appear to
  signal the presence of a stable/unstable manifold of the \ac{NHIM} that
  lies between them.\cite{Mancho2013}
  Although there are known examples that the singular contour of the \ac{LD}
  correctly corresponds to the stable/unstable manifolds,
  there is no general proof on the correspondence.
  Formally, the \acp{LD} would be obtained at $t\to\infty$ where the values 
  all go to infinity regardless of the choice of trajectory. 
  The exceptions arise from fixed points at which the \ac{LD} 
  of trajectories asymptotic to them
  converges to a finite value. 
  In practice, we integrate for a long, but finite, time at which there 
  is a visible feature,
  deviation, or
  ``abrupt change''
  in the \ac{LD}
  between initially nearby trajectories,
  or abrupt features in the \ac{LD} such as narrow ridges or valleys.

  There are some concerns about the generality of the conjecture.
  In particular, Haller\cite{haller16a} criticized the use of the \ac{LD}
  because it is not objective. 
  That is, as long as the \ac{LD} is defined by a norm, such as an arc length,
  the \ac{LD} value is not necessarily independent of
  the particular choice of \acp{DoF}.
  To illustrate this concern,
  let us consider a three-dimensional normal mode Hamiltonian
  which has one reactive \ac{DoF} and two vibrational \acp{DoF}.
  If the trajectory is on the \ac{NHIM},
  the dynamical variables $(q, p)$ of the reactive \ac{DoF} remain constant,
  {\it i.e.}, they remain on the saddle with zero reactive velocity.
  If \ac{LD} values are uniform over the two vibrational \acp{DoF},
  the reactive \ac{DoF} is then the only relevant \ac{DoF}.
  However,
  this is not the case as shown in Fig.~\ref{fig:NM_LD}.
  On the other hand,
  for the reactive mode $(q_0,p_0)$ of a normal mode Hamiltonian,
  a trajectory on the stable and unstable manifolds of the \ac{NHIM},
  with a momentum
  expressed
  by $p_0(0)\ee^{-\lambda t}$ and $p_0(0)\ee^{\lambda t}$,
  reaches the \ac{NHIM} at $t\rightarrow \infty$ and $t\rightarrow -\infty$ respectively.
  This means that around the \ac{NHIM}, extremal \ac{LD}
  can be affected by the reactive \ac{DoF} and
  be prone to dominant contributions from the vibrational \acp{DoF}.
  In some cases,
  modification of some initial conditions along the manifold
  can have larger effects on the \ac{LD} value than from those
  not along the manifold.
  For these cases, the `abrupt' change in the \ac{LD} value could
  misidentify the region as containing \iac{NHIM}\@.
  Thus the use of the \ac{LD} to identify \acp{NHIM} has to be done with care.

\subsection{Asymptotic Trajectory Indicator}\label{ssec:ATI_Theory}

  \begin{figure}[t]
    \includegraphics[clip=true,width =\figurewide]{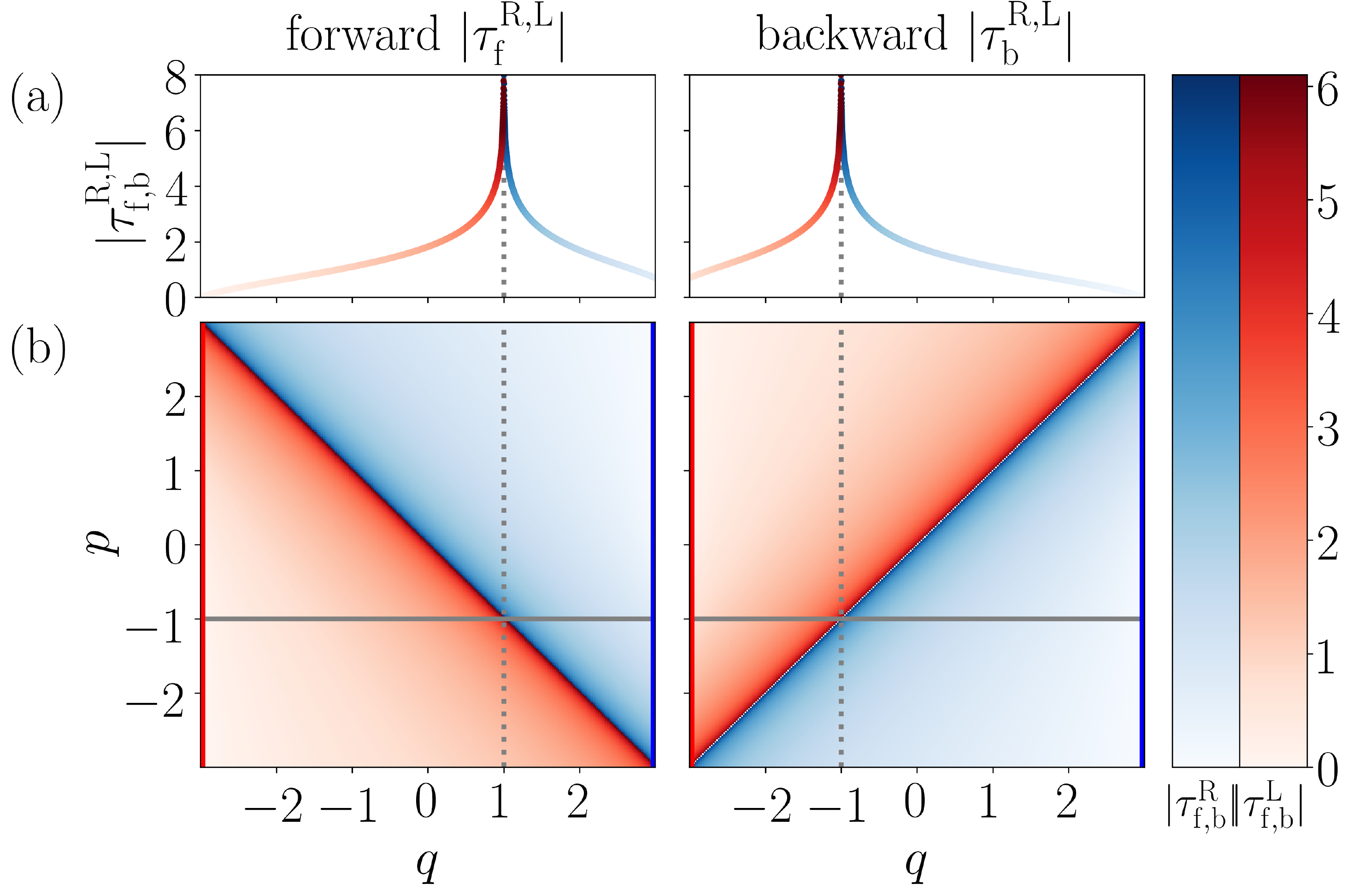}
    \caption{%
      The forward \ac*{ATI},
      $\tau_\mathrm{f}^\mathrm{R,L}$ (left column),
      and backward \ac*{ATI},
      $\tau_\mathrm{b}^\mathrm{R,L}$ (right column), 
      defined in Eqs.~\eqref{eq:ATIf}
      and~\eqref{eq:ATIb} respectively.
      They correspond respectively to the first hitting time of 
      the right $q=3$ (blue) and left $q=-3$ (red)
      absorbing boundary starting from the given point.
      Panel (a) shows the values along the gray line plotted in phase space (b).
      The dashed line corresponds to the singular point along the gray line in (b).
      The Hamiltonian is the same as in Fig.~\ref{fig:HyperbolicFlow}.
    }\label{fig:NM_ATI}
  \end{figure}

  An alternative to the \ac{LD} can be achieved through the
  reactivity bands (in one-dimensional domains) 
  or reactivity map (in higher dimension)\cite{Wall1958,Wall1961,Wall1963,
  Wright1975,Wright1976,Wright1977,Wright1978,Laidler1977,Tan1977}
  and reactivity boundary.\cite{komatsuzaki13a}
  These structures are defined on the domain of initial
  condition in phase space
  by designating them according to their ultimate 
  origin or destination 
  to a reactant domain
  or one of possibly many distinct product domains.
  Between initial conditions assigned to different final basins,
  there could be initial conditions whose trajectories
  never reach one of the designated basins, and
  thus act as reactivity boundary.\cite{komatsuzaki13a}
  For example,
  the purple and pink trajectories in Fig.~\ref{fig:HyperbolicFlow}
  form the boundary between initial conditions
  assigned to products (at $t\to+\infty$)
  and reactants (at $t\to-\infty$), respectively.
  Together,
  these boundaries separate trajectories into four categories
  whose initial and final chemical state
  can be assigned according to whether
  trajectories are
  (1) staying inside of, (2) entering into,
  (3) exiting from, or (4) staying outside of the trapping area.
  Those structures are fundamental to 
  the turnstile\cite{Mackay1984}
  and reaction island\cite{deLeon1} theories.

  To compare with
  the \ac{NHIM} theory,\cite{Fenichel72,Wiggins1994,Eldering2013}
  let us formulate the reactivity boundaries mathematically
  by using their asymptotic nature.
  Let $\bm{\phi}^t$ be the propagator in time $t$,
  {\it i.e.}, a {\it flow function},
  \begin{equation}
    \bm{\phi}^t:\bm{x}(t_0)\rightarrow \bm{x}(t_0+t).
  \end{equation}
  For given $\bm{\phi}^t$, there could be a set of trajectories,
  which is restricted to a subspace $\mathcal{M}$.
  {Such} a subspace $\mathcal{M}$ of the phase space $\mathcal{P}$
  is said to be {\it invariant\/} when
  \begin{equation}
    \mathcal{M}:= \left\{\bm{x}\middle|
    \forall\bm{x}\in\mathcal{M}~\mathrm{and}~\forall t,
    \bm{\phi}^t(\bm{x})\in\mathcal{M} \right\}.
  \end{equation}
  Or more simply,
  $\mathcal{M}$ is an invariant subspace when
  $\bm{\phi}^t(\mathcal{M})=\mathcal{M}$ for arbitrary $t$.
  If there are asymptotic trajectories to $\mathcal{M}$
  then one can define the {\it stable manifold\/} and {\it unstable manifold\/}
  of the invariant manifold $\mathcal{M}$
  as follows:
  \begin{eqnarray}
    \mathcal{W}_\mathcal{M}^\mathrm{(s)}:=&\left\{\bm{x}\middle|
    \forall\bm{x}\in\mathcal{P},
    \lim_{t\rightarrow \infty}\bm{\phi}^t(\bm{x})\in\mathcal{M}\right\},
    \label{eq:stableM}\\
    \mathcal{W}_\mathcal{M}^\mathrm{(u)}:=&\left\{\bm{x}\middle|
    \forall\bm{x}\in\mathcal{P},
    \lim_{t\rightarrow-\infty}\bm{\phi}^t(\bm{x})\in\mathcal{M}\right\}.
    \label{eq:unstableM}
  \end{eqnarray}
  Accordingly,
  $\mathcal{W}_\mathcal{M}^\mathrm{(s)}\cap \mathcal{W}_\mathcal{M}^\mathrm{(u)}$
  is invariant.
  For simplicity, we define the reactivity boundary separating
  the destination and origin of trajectories as
  $\mathcal{W}_\mathrm{asym}^\mathrm{(s)}$
  and
  $\mathcal{W}_\mathrm{asym}^\mathrm{(u)}$
  respectively.
  This allows us to designate an invariant manifold
  related only to
  $\mathcal{W}_\mathrm{asym}^\mathrm{(s)}$ and
  $\mathcal{W}_\mathrm{asym}^\mathrm{(u)}$ as
  \begin{equation}\label{eq:asymM}
    \mathcal{M}_\mathrm{asym}:=
    \mathcal{W}_\mathrm{asym}^\mathrm{(s)}
    \cap\mathcal{W}_\mathrm{asym}^\mathrm{(u)},
  \end{equation}
  where
  $\mathcal{W}_X^\mathrm{(s)}
  :=\mathcal{W}_{\mathcal{M}_X}^\mathrm{(s)}$, and
  $\mathcal{W}_X^\mathrm{(u)}
  :=\mathcal{W}_{\mathcal{M}_X}^\mathrm{(u)}$
  for $X=\mathrm{asym}$.

  In practice, 
  this manifold $\mathcal{M}_\mathrm{asym}$
  can be detected through observation of nearby trajectories.
  We first identify a subspace
  $\mathcal{S}$ that contains $\mathcal{M}_\mathrm{asym}$
  ---{\it i.e.}, $\mathcal{S}$ $\supset\mathcal{M}_\mathrm{asym}$,---
  with initial conditions associated with trajectories that 
  remain in $\mathcal{W}_\mathrm{asym}^\mathrm{(s)}$
  under forward propagation or
  $\mathcal{W}_\mathrm{asym}^\mathrm{(u)}$ 
  under backward propagation.
  We then define the {\it first passage time\/} for each point $x\in\mathcal{S}$
  according to when it first crosses the boundary $\partial\mathcal{S}$,
  that is,
  \begin{equation}\label{eq:ATIf}
    \tau_\mathrm{f}(\bm{x};\mathcal{S}):=
    \min_t \left\{t\middle|
    \forall t\geq 0, \bm{\phi}^t(\bm{x})\in\partial\mathcal{S}\right\}.
  \end{equation}
  Consequently, for $\bm{x}\in\mathcal{W}_\mathrm{asym}^\mathrm{(s)}$,
  then $\tau_\mathrm{f}(\bm{x};\mathcal{S})=\infty$.
  Neighbors $\bm{y}$ 
  of
  $\bm{x}\in \mathcal{W}_\mathrm{asym}^\mathrm{(s)}$,
  will necessarily have large but finite $\tau_\mathrm{f}(\bm{y};\mathcal{S})$
  because of
  the continuity of the equation of motion.
  Therefore,
  abrupt changes in $\tau_\mathrm{f}(\bm{x};\mathcal{S})$
  indicate the nearby location of the {\it singular contour}, {\it i.e.},
  $\mathcal{W}_\mathrm{sing}^\mathrm{f}:=\left\{\bm{x}\middle|
  \tau_\mathrm{f}(\bm{x};\mathcal{S})=\infty\right\}$.
  This contour is formed from forward asymptotic trajectories,
  and is therefore the manifold $\mathcal{W}_\mathrm{asym}^{(s)}$.
  For this reason, hereafter we call $\tau_\mathrm{f}(\bm{x};\mathcal{S})$
  the {\it forward asymptotic trajectory indicator\/} (ATI).
  Similarly, we can define
  \begin{equation}\label{eq:ATIb}
    \tau_\mathrm{b}(\bm{x};\mathcal{S}):=
    \max_t \left\{t\middle|
    \forall t\leq 0, \bm{\phi}^t(\bm{x})\in\partial\mathcal{S}\right\}
  \end{equation}
  as the {\it backward \ac{ATI}\/}\@.
  We provide an example in Fig.~\ref{fig:NM_ATI} to show a typical behavior of the \ac{ATI}.

  It is useful to consider how the \ac{ATI} generalizes for more complex cases,
  including those when $\partial S$ encloses multiple asymptotic manifolds.
  In the following, we prove that
  when one observes an ATI value of a point on $\partial S$,
  its value is {\it almost surely\/} finite, if $S$ is compact.
  Here, ``almost surely'' means that the statement is true
  except for one or more dimensions less than an equi-energy surface of the phase space.
  % This finiteness of ATI can be proved by following.
  First, let us consider a subset of initial conditions on $\partial S$,
  for which entire trajectories are not bounded to the region $S$.
  Each such trajectory must have an even number of intersections on $\partial S$
  ---that is, they go in and out in pairs---
  because it is not bounded.
  Second, the other initial conditions are known to be almost surely recurrent
  as specified by {\it Poincar{\'e}'s recurrence theorem}.
  (See, {\it e.g.}, Sec.~3 16 D in Ref.~\onlinecite{Arnold1989}.)
  Asymptotic trajectories serves as examples of such measure zero sets.
  Therefore,
  when one observes an ATI value of a point on $\partial S$,
  its value is almost surely finite, if $S$ is compact.
  In fact, below 
  % in Figs.~\ref{fig:global}, % Better to not be too specific about which figure becasue
  % if we do that, we have to renumber figures, and discuss them here...
  we observe the reactivity boundaries as singular contours
  even if $S$ includes multiple asymptotic manifolds.
  In practice,
  the finiteness of the \ac{ATI} may indicate the requirement of very long time propagation.
  For example, a trajectory may be trapped in a potential energy well.
  The complexity of highly coupled reactions also challenges our approach
  because they can lead to fractal structures arising from chaotic dynamics.
  Such cases can be avoided by taking $S$ around a $\mathcal{M}_\mathrm{asym}$.
  In Subsec.~\ref{ssec:ATI},
  we further show heuristics to avoid long-time trajectory calculations.

\subsection{Comparison with \ac{NHIM} theory}
  Here we confirm that the \ac{ATI} leads to \iac{NHIM} in cases when 
  the solution is accessible
  to perturbation theory, and how it extends beyond it.
  To this end, we first observe that the \ac{NHIM} Persistence Theorem
  provides for the 
  persistence of the phase space geometry ---vis-{\`a}-vis the \ac{NHIM}---
  under perturbation.
  As we reconfirm in the Appendix~\ref{apx:NHIMtheorem},
  the reaction dynamics
  around an index-one saddle on a potential energy surface
  satisfies the requisite conditions of the theorem.
  In this case,
  the \ac{NHIM} corresponds to
  a \acp{DoF} orthogonal to the reactive \ac{DoF} at the \ac{TS}.
  The latter is the nonlinear analogue of
  the \ac{DoF} associated with the eigenvector of the positive Hessian eigenvalue.

  The stable and unstable manifolds associated to
  the reaction coordinate
  are illustrated in Fig.~\ref{fig:HyperbolicFlow}.
  They are associated with imaginary frequencies,
  called characteristic exponents throughout this work,
  that characterize their strongest decay.
  That is, the persistence of the manifolds results from their
  exponential expansion and contraction, respectively, and
  the sign of the characteristic exponent reflects this.
  Suppose for
  its imaginary frequency $\lambda_\mathrm{s,u}$
  and upper bound of the other imaginary part of frequencies $\lambda_\mathrm{c}$,
  the \ac{NHIM} has $r\ge 1$ such that,
  $0\le r\lambda_\mathrm{c}<\lambda_\mathrm{s,u}$
  ({\it e.g.}\ $\lambda_\mathrm{c}=0$ when the saddle is index one).
  In this case and if the flow is $k$-differentiable $\bm{\phi}^t\in C^k$,
  then $\mathcal{M}_\mathrm{NHIM}\in C^k$ when $k\le r$.
  In addition,
  for a $f\in C^k$,
  $f(\mathcal{M}_\mathrm{NHIM})$ is still \iac{NHIM} and $k$-differentiable,
  {\it i.e.}, persistent under $C^k$ perturbation.
  Specifically, this \ac{NHIM} is called a $k$-NHIM.\cite{Hirsch1977,
  Wiggins1994,Eldering2013}
  (See Appendix~\ref{apx:NHIMtheorem} for the explicit definition).

  For example, if $\mathcal{M}(\subset\mathcal{S})$ is \iac{NHIM}
  associated with an index-one saddle,
  the characteristic exponents $-\lambda_\mathrm{s,u}$ and $\lambda_\mathrm{s,u}$
  corresponding to expansion and contraction directions
  are associated with stable $\mathcal{W}_\mathrm{NHIM}^\mathrm{(s)}$ and
  unstable $\mathcal{W}_\mathrm{NHIM}^\mathrm{(u)}$ manifolds, respectively.
  The singular contour of the \acp{ATI}
  defined in the previous section
  is a stable 
  ({\it i.e.}, $\mathcal{W}_\mathrm{sing}^\mathrm{f}$)
  or unstable 
  ({\it i.e.}, $\mathcal{W}_\mathrm{sing}^\mathrm{b}$)
  manifold of the \ac{NHIM}\@.

  More generally, 
  $\mathcal{M}_\mathrm{NHIM}$ and $\mathcal{M}_\mathrm{asym}$
  are not equivalent for higher index saddles.
  For example,
  if there are other exponentially growing directions with characteristic exponents
  $-\mu_\mathrm{s}$ and $\mu_\mathrm{u}$
  in addition to those of
  that stable $-\lambda_\mathrm{s}$ and unstable $\lambda_\mathrm{u}$ manifolds,
  satisfying
  $-\lambda_\mathrm{s}<-\mu_\mathrm{s}<0<\mu_\mathrm{u}<\lambda_\mathrm{u}$,
  then the directions are tangent to $\mathcal{M}_\mathrm{NHIM}$
  because of $\lambda_\mathrm{c}\ge \max\{\mu_\mathrm{s},\mu_\mathrm{u}\}$.
  On the other hand, those directions are still normal to $\mathcal{M}_\mathrm{asym}$
  because they converge to it asymptotically.
  Therefore,
  $\mathcal{M}_\mathrm{asym}$ is not \iac{NHIM},
  and $\mathcal{M}_\mathrm{asym}$ may not have persistence
  under perturbation.
  Such a failure was recently seen for an index-2 saddle 
  despite the convergence of perturbation theory.\cite{Nagahata2013a}

  The reactivity boundary
  ($\mathcal{W}_\mathrm{asym}^{(x)}$ with $x=\mathrm{s,u}$)
  is relevant to reaction dynamics
  because it marks the transition at which the fate of the 
  reactants is cast.
  For example, for a reaction with an index-1 saddle,
  the reactivity boundaries are the same as the
  stable and unstable manifolds of the \ac{NHIM}\@.
  This feature allows us to detect the boundary of reactivity
  even for reactions
  associated with higher index saddles.\cite{komatsuzaki13a}

	\subsection{\ac*{NHIM} of Random Dynamical Systems}\label{ssec:RDS}

  The nature of a trajectory resulting from \iac{SDE}
  is completely different from an \ac{ODE}.\cite{Oksendal2003,Duan2015}
  For example, a particle whose motion is described by 
  a Langevin equation 
  is thereby driven stochastically by a random force $\xi(t)$
  only in a formal sense.
  In practice, one must integrate over the latter to
  propagate the particle within a difference equation,
  {\it e.g.}, by the Euler-Maruyama method.\cite{Kloeden1992}
  Integrals of the random force
  are neither smooth nor continuous in the usual sense.
  Its statistics 
  generally satisfy that of a Wiener process.\cite{Duan2015}
  Specifically, we can define the difference in the 
  accumulated random force across a time interval as
  \begin{equation}
    B_{s+t}-B_s \equiv
    \int_s^{s+t} \xi(\tau)\dd \tau
    \sim \mathcal{N}(0,Dt),
    \label{eq:RF}
  \end{equation}
  where $D$ is the diffusion constant.
  The relation $X\sim \mathcal{N}(\mu,\sigma^2)$ is defined such that 
  $X$ is a random variable resulting from a normal distribution
  with mean $\mu$ and variance $\sigma^2$.
  With this construction, $B_t$
  has continuity in a weak sense;
  that is, it satisfies
  the $\beta$-H{\"{o}}lder continuity 
  for all $\beta$ satisfying $\beta<1/2$.\cite{Oksendal2003}
  Recall that  
  $\beta$-H{\"{o}}lder continuity requires the
  existence of a constant $c>0$ and
  an exponent $\beta>0$
  such that 
  \begin{equation}
    |B_{t}-B_{s}|\le c |t-s|^\beta.
  \end{equation}
  for all times $t$ and $s$.
  We further require $\beta=1$ so that the function is differentiable.
  In the present case,
  the existence of weak continuity allows us to 
  claim 
  uniqueness of the solution for a given stochastic sequence,
  and hence
  each solution has pathwise uniqueness.\cite{Oksendal2003}
  Second, as a consequence of the lack of smoothness
  in $B_t$,
  it is nowhere differentiable along $t$,
  and it can not be inverted
  to a $\xi$.

  The \ac{NHIM} Persistence Theorem describing the geometry of the solutions of the \ac{SDE} 
  can be framed through an analysis of the 
  paths (trajectories) of the \ac{SDE}\@.
  For each $d$-dimensional accumulated random force $\bm{B}_t$,
  one can uniquely obtain 
  a $d$-dimensional, $t$-continuous function $\bm{\omega}(t)$
  that is associated with the saddle point
  (precisely defined in Appendix~\ref{apx:RDEfromSDE})
  at each instance of time $t$, and for which we are free to 
  initialize at $\bm{\omega}(0)=0$.
  (Note that this $\bm{\omega}(t)$ is simply an abstraction of the so-called
  \ac{TS} trajectory.\cite{dawn05a})
  Then a probability is determined from a bundle of instances $\bm{\omega}$.
  Let us also introduce a $t$-origin shift to the probability measure,
  the so-called {\it Wiener shift\/} $\theta_s$, such that,
  \begin{equation}
    (\theta_{s} \bm{\omega})(t)=\bm{\omega}(s+t)-\bm{\omega}(s),~s>0.
  \end{equation}
  This $\theta_s$ introduces a shift in the initial time of a stochastic process
  to $s$.
  For example, any given Brownian motion can be written as
  a particular manifestation of $\bm{\omega}$, such that,
  for $\bm{B}_t=\bm{B}[\bm{\omega}(t)]$,
  \begin{eqnarray}
    \bm{B}[(\theta_s \bm{\omega})(t)]
    =&\bm{B}[\bm{\omega}(s+t)-\bm{\omega}(s)]\nonumber\\
    =&\bm{B}[\bm{\omega}(t+s)]-\bm{B}[\bm{\omega}(s)].
  \end{eqnarray}
  The relation can be found in Eq.~(6.46) of Ref.~\onlinecite{Duan2015}.
  (See also Eq.~\eqref{eq:RF} for \iac{SDE}.)
  We can then define stochastic analogues of the flow function
  and invariant manifold associated with an \ac{ODE} 
  to the analogues of the \iac{SDE}:
  the stochastic {\it cocycle\/} $\bm{\phi}^t_{\bm{\omega}}$,
  \begin{equation}
    \bm{\phi^t_{\bm{\omega}}}:
    \bm{x}(s)\rightarrow\bm{x}(t+s;\theta_{s}\bm{\omega},\bm{x}(s)),
    \label{eq:RDS}
  \end{equation}
  and the {\it random invariant manifold\/} $\mathcal{M}(\bm{\omega})$,
  \begin{equation}
    \bm{\phi}^t_{\bm{\omega}}(\mathcal{M}(\bm{\omega}))
    =\mathcal{M}(\theta_t\bm{\omega}),\label{eq:RIM}
  \end{equation}
  respectively.
  In Eq.~\eqref{eq:RDS}, the stochastic path $\bm{x}(\cdot)$ starts at $s$
  requiring a shift
  in the time origin of $\bm{\omega}(t)$ to $s$,
  and hence the need for the term $\theta_s\bm{\omega}$ in the argument of $x$.
  \Iac{RDS} defined by Eq.~\eqref{eq:RDS} is uniquely obtained from
  a {\it \ac{RDE}\/} with the vector field $\bm{f}$
  \begin{equation}
    \dot{\bm{x}}=\bm{f}(\theta_t\bm{\omega},\bm{x}),
    \label{eq:RDE}
  \end{equation}
  as usually obtained for \ac{ODE}.
  One can obtain this \ac{RDS} for some classes of \acp{SDE}, 
  including Langevin type equations.
  (See Appendix~\ref{apx:RDEfromSDE}.)

  The \ac{NHIM} Persistence Theorem for a \ac{RDS} holds\cite{Li2013}
  for random invariant manifolds defined by Eq.~\eqref{eq:RIM}.
  The remarkable result of the theorem is that
  the \ac{NHIM} is persistent under $C^1$ perturbations
  and is $C^k$-smooth at time $t$ when $\bm{\phi}^t_{\bm{\omega}}\in C^k$.
  The smoothness of $\bm{\phi}^t_{\bm{\omega}}$ does not indicate
  that the variables of \ac{SDE}
  ---{\it e.g.}\ of the Langevin equation--- are smooth
  over integration time.
  On the other hand, \ac{SDE} still has a H{\"{o}}lder continuity as 
  detailed above.
  Because of this continuity,
  the closer an initial condition is
  to the stable or unstable manifold of \iac{NHIM}, then 
  the longer the trajectory will spend in time around the \ac{NHIM}.
  In this way,
  there can still exist abrupt changes and singular contours
  in the \acp{ATI} of \iac{SDE},
  where the latter corresponds to the
  \ac{NHIM} and its stable and unstable manifolds.
  Because of the theorem, these manifolds are smooth.

  As was suggested in Refs.~\onlinecite{Li2013,Eldering2013},
  the NHIM Persistence Theorem holds for autonomous systems.
  For such systems,
  there is a single $\bm{\omega}$
  which corresponds to the time series of the external force\cite{Arnold1998rds}
  and does not require any fundamental change
  to satisfy the conditions needed to satisfy the
  \ac{NHIM} Persistence Theorem.\cite{Eldering2013}

\section{Neighbor Bisection and Continuation with ATI}\label{sec:Method}

  Thus far,
  we have considered the case of a single invariant manifold
  $\mathcal{M}_{\mathrm{asym}}$ defined in Eq.~\eqref{eq:asymM}
  that lives within the subspace $\mathcal{S}$ of the phase space $\mathcal{P}$
  without considering the computational requirements for its implementation.
  In practice, the latter is exacerbated by the
  existence of multiple \acp{NHIM} in $\mathcal{S}$.
  Nevertheless, typically some features of its structure 
  (such as possible location or locations) are 
  approximately known, and they
  can be used to optimize sampling of candidate points of the manifold,
  and improve the numerical implementation of the search.
  Here,
  we present a practical approach for defining the absorbing boundaries,
  sampling the surrounding neighborhood of the reactivity boundary, and
  visualizing the \ac{ATI}.
  We introduce the \ac{NBC-ATI} method to effectively increase
  the resolution and accuracy of the reactivity boundaries.

\subsection{Absorption and Visualization}

  \begin{figure}[t]
    \includegraphics[clip=true,width =\figurewide]{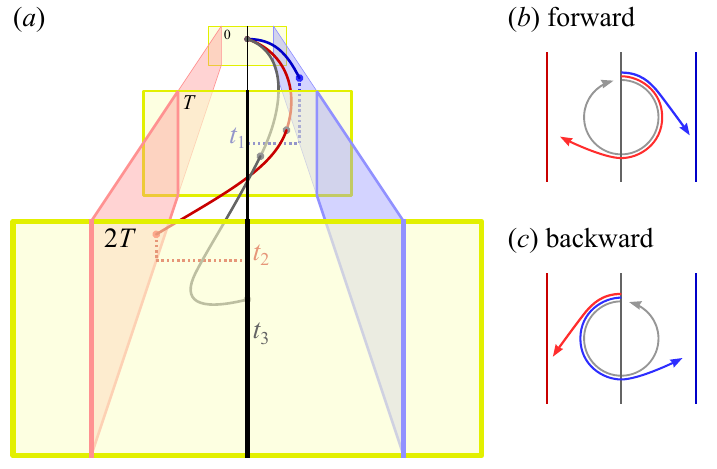}
    \caption{%
      A schema of possible trajectories from an initial point
      in the region relative to the absorbing boundary surfaces is shown in 
      panel (a) as a function of time:
      A trajectory hitting the left (in red) or right (in blue) absorbing boundary,
      and 
      a recurrent trajectory (in gray) which returns to the same coordinate
      with the same sign in the velocity.
      The projection of these trajectories
      in forward and backward time onto the phase space 
      is shown in panels (b) and (c), respectively.
    }\label{fig:abs}
  \end{figure}

  The values of 
  the forward $\tau_\mathrm{f}$ and backward $\tau_\mathrm{b}$ times
  defined in Sec.~\ref{sec:Theory}
  depend on the initial and absorbing conditions.
  However, the location 
  of the singular contour in $\tau$
  corresponding to the reactivity boundary
  is independent of the absorbing conditions,
  as we described in Sec.~\ref{ssec:ATI_Theory}.
  A naive absorbing condition can be
  defined through a coordinate $\bm{q}$ at a value that is 
  significant to the dynamics
  ({\it e.g.}, at a potential energy minimum)
  and sufficiently far away from the \ac{TS}.
  Figure~\ref{fig:abs} illustrates an example in which
  trajectories are absorbed 
  and thereafter assigned a first hitting time $\tau$
  ---viz, the \ac{ATI}.

  The domain of initial conditions can be 
  labeled using a color that denotes 
  one of the surfaces shown in Fig.~\ref{fig:abs}
  to which they absorb, and
  an intensity commensurate with the value of the 
  \ac{ATI}, $\tau$.
  The result for a normal mode Hamiltonian 
  is illustrated in Fig.~\ref{fig:NM_ATI}.
  Initial conditions on the domain 
  are labeled in red or blue 
  according to whether 
  trajectories are absorbed on the 
  left at $q=-3$ or right at $q=3$, respectively.
  The intensity of the color is
  commensurate with the value of the \ac{ATI},
  $\tau_\mathrm{f}^\mathrm{L}$ or $\tau_\mathrm{f}^\mathrm{R}$.
  In the context of the theory described in Sec.~\ref{sec:Theory},
  the absorbing boundaries at $\mathrm{R}$ and $\mathrm{L}$ employed
  here are examples
  of the two disconnected subsets of $\partial\mathcal{S}$.

  As shown in Fig.~\ref{fig:NM_ATI}a,
  $\tau_\mathrm{f}$ or $\tau_\mathrm{b}$ has a singular contour
  for the forward or backward asymptotic trajectories
  that corresponds to the purple or pink trajectories
  in Fig.~\ref{fig:HyperbolicFlow}, respectively.
  Hence, the abrupt change in $\tau_\mathrm{f}$ or $\tau_\mathrm{b}$
  in Fig.~\ref{fig:NM_ATI}b indicates that
  there is, indeed, a reactivity boundary nearby.
  This singular contour
  is the stable or unstable manifold of the \ac{NHIM}.
  For the present case, the \ac{NHIM}
  is $(q,p)=(0,0)$,
  the stable manifold is $p=-q$,
  and the unstable manifold is $p=+q$.

\subsection{Efficient Sampling Algorithm}\label{ssec:Algorithm}

  \begin{figure*}[t]
    \includegraphics[clip=true,width =\figurewide]{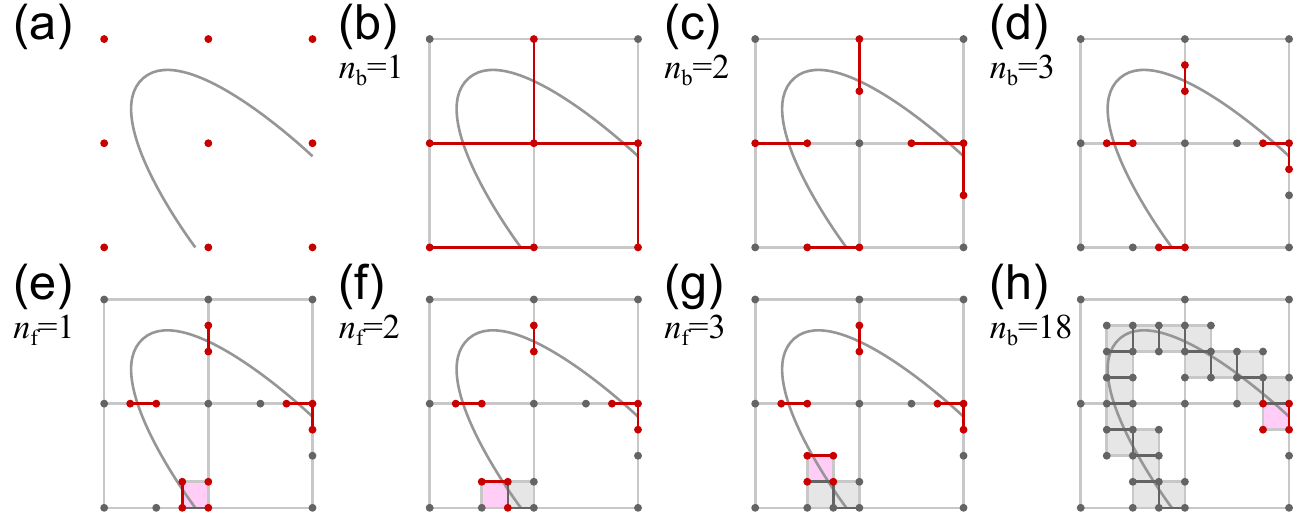}
    \caption{%
      Schema of selected steps in the sampling algorithm as described in the text.
      In any given step, the active straddling pairs (red), points (red) and area (magenta),
      are highlighted in color, and previously sampled
      straddling pairs (gray), area (light gray) and point (gray) 
      are highlighted in gray scale.
    }\label{fig:Algorithm}
  \end{figure*}

  When analytical methods, such as perturbation theory, fail to 
  produce the exact form of the reactivity boundary, we must resort
  to using numerical methods.
  The challenge to the numerics, however, 
  lies in the fact that the reactivity boundary 
  has measure zero, and hence statistical sampling is inefficient.
  To overcome this challenge, we employ an algorithm
  which effectively generalizes the one-dimensional bisection
  method to 2D and the required higher dimensionality of the space that contains the
  reactivity boundary.
  It may be implemented 
  iteratively when a user needs to improve resolution
  to, for example, increase the number of 
  points of the reactivity boundary.

  To initiate the first stage of the algorithm, we need 
  a low-resolution representation of the reactivity boundary.
  We can construct this by way of creating a low-resolution grid in the 
  domain, and performing a ``brute-force search'' across all the
  vertices to identify pairs of points that straddle the reactivity boundary
  in the sense that one of the two adjacent points goes 
  to the reactant-side surface and the other to the product-side surface.
  We call such pairs of points {\it straddling pair\/}s.
  In more detail, 
  we execute the brute-force search on a $(2^{N_\mathrm{BF}}+1)\times(2^{N_\mathrm{BF}}+1)$ grid
  ($N_\mathrm{BF}=1$ in Fig.~\ref{fig:Algorithm}a).
  The grid results in a spacing resolution at $\bm{L}/2^{N_\mathrm{BF}}$
  where for the length of given 2D rectangle domain $\bm{L}=(L_x,L_y)$.
  As long as the reactivity boundary does not fold within a width of this
  resolution, a straddling pair will capture one of its points within its
  connecting segment.

  In the first stage --- ``seed refinement'' ---,
  we refine the initial straddling pairs
   up to the desired resolution.
  Specifically, we iterate the bisection method $N_\mathrm{res}-N_\mathrm{BF}$ times
  to find a new set of straddling pairs to the desired grid resolution,
  {\it i.e.}, $(2^{N_\mathrm{res}}+1)\times(2^{N_\mathrm{res}}+1)$.
  In this stage, we assume that
  the desired structures are all larger than the resolution of the initial grid.
  The resulting resolution of the straddling pairs,
  illustrated in Fig.~\ref{fig:Algorithm}b-d for $N_\mathrm{res}=3$,
  is on the order of $\bm{L}/2^{N_\mathrm{res}}$.

  In the second stage --- ``path-following'' or ``continuation'' ---,
  we construct edge-to-edge and recurrent chains of straddling pairs.
  To construct edge-to-edge chains,
  we first choose one of straddling pairs
  which are on the edge of the domain.
  If it exists,
  then we use the pair to identify a next pair
  (red line in Fig.~\ref{fig:Algorithm}e-h)
  that also straddles the reactivity boundary.
  A chain of straddling pairs is constructed by repeating this iteration
  until it reaches the edge of the domain (Fig.~\ref{fig:Algorithm}h).
  If after this construction, there remains straddling points on the edge,
  we pick one of them and apply the path-following stage above until they are exhausted.
  To construct recurrent chains,
  we then repeat the procedure on remaining straddling pairs inside the domain,
  which will necessarily end on themselves,
  until all straddling pairs are exhausted.

  In the third stage --- ``precision refinement'' ---,
  we refine the precision of the straddling pairs of the obtained chains.
  We apply the bisection method to each identified straddling pair
  $N_\mathrm{pre}-N_\mathrm{res}$ times
  to achieve the desired grid precision $(2^{N_\mathrm{pre}}+1)\times(2^{N_\mathrm{pre}}+1)$
  with $\bm{L}/2^{N_\mathrm{pre}}$.
  In the numerical applications, we chose $N_\mathrm{pre}=30$.
  Because $2^{-30}\approx 10^{-9}$ and the highest resolution in double-precision is
  16 decimal digits, 
  then the points between straddling pairs can be differentiated 
  only up to $7$ additional digits.
  If we use ${10}^{-13}$ for the desired accuracy in the time integration,
  then we still have $4$ reliable such digits in a single iteration of time propagation.
  Thus, for this setup, the computational result is reliable
  as long as the accumulated numerical error through the time propagation
  for obtaining $\tau$ is less than
  a factor of $10^{4}$ times the error of a single step iteration.

  \begin{table*}[t]
    \caption{%
      Number of evaluated trajectories in each stage of the sampling algorithm.
      $C_\mathrm{st}[resolution;\mathcal{M}]$ and $C_\mathrm{sq}[resolution;\mathcal{M}]$
      are the number from straddling pairs and that from covering squares of $\mathcal{M}$
      in the given $resolution$, respectively.
    }
    \begin{tabular}{llll}
      \toprule
      Stage & Method & \multicolumn{2}{c}{Number of evaluated trajectories}\\
      & & Exact & Order (expected) \\
      \midrule
      0th & Brute-Force & ${(2^{N_\mathrm{BF}}+1)}^2$ & $2^{2N_\mathrm{BF}}$ \\
      1st & Bisections & $C_\mathrm{st}{[2^{N_\mathrm{BF}}+1;\mathcal{M}]}\times (N_\mathrm{res}-N_\mathrm{BF})$
          & $2^{N_\mathrm{BF}}\times (N_\mathrm{res}-N_\mathrm{BF})$ \\
      2nd & Path-Following & $C_\mathrm{sq}{[2^{N_\mathrm{res}}+1;\mathcal{M}]}-C_\mathrm{st}{[2^{N_\mathrm{BF}}+1;\mathcal{M}]}$
          & $2^{N_\mathrm{res}}$ \\
      3rd & Bisections & $C_\mathrm{st}{[2^{N_\mathrm{res}}+1;\mathcal{M}]}\times (N_\mathrm{pre}-N_\mathrm{res})$
          & $2^{N_\mathrm{res}}\times (N_\mathrm{pre}-N_\mathrm{res})$ \\
      \bottomrule
    \end{tabular}\label{tbl:order}
  \end{table*}

  Suppose that we use a $(2^{N_\mathrm{BF}}+1)\times(2^{N_\mathrm{BF}}+1)$ grid
  in defining our initial search space,
  and we wanted to get a resolution of
  $\delta \bm{L}/2^{N_\mathrm{res}}$.
  Naively, this would require the determination of ${(2^{N_\mathrm{res}}+1)}^2$
  points on a two-dimensional grid.
  Using our algorithm, instead,
  we expect that the number of points is
  approximately proportional to $2^{N_\mathrm{res}}$.
  This estimate is based on an assumption
  that the number of straddling pairs on a
  $(2^{N_\mathrm{res}}+1)\times (2^{N_\mathrm{res}}+1)$ grid
  $C_\mathrm{st}{[2^{N_\mathrm{res}}+1;\mathcal{M}]}$ is proportional to
  $2^{N_\mathrm{res}}$,
  where $\mathcal{M}$ is the manifold whose straddling pairs we are identifying.
  The assumption is trivially correct when $\mathcal{M}$ is one-dimensional
  on the observed two-dimensional domain.
  In that case, the order estimation can be made with the relation:
  $C_\mathrm{st}[x;\mathcal{M}]\le C_\mathrm{sq}[x;\mathcal{M}] \le (9/5)C_\mathrm{st}[x;\mathcal{M}]$,
  where
  $C_\mathrm{sq}$ is the number of points attach to the squares with resolution $x$
  covering $\mathcal{M}$, and
  the first and the second equality are from the cases
  when all the straddling pairs have a different and the same
  orientation from that of the adjacent pair, respectively.
  We know that the
  $N$-fold application of the bisection method to a straddling pair 
  needs the calculation on $N$ additional points,
  and the number of input seeds for the path-following  stage
  is $C_\mathrm{st}{[2^{N_\mathrm{BF}}+1;\mathcal{M}]}$.
  We listed the order of magnitude
  for the number of points for each stage in Table~\ref{tbl:order}.
  Based on this table,
  if we chose $2^{N_\mathrm{BF}}\ll 2^{N_\mathrm{res}}$,
  then the order can be estimated as $2^{N_\mathrm{res}}$.
  When we include the third stage in the estimation,
  the order becomes $N_\mathrm{pre}-N_\mathrm{res}+1$ times larger
  than the order up to the second step.

  For $n$ \ac{DoF} systems at a constrained energy,
  the number of 2D slices required
  to visualize the entire reactivity boundary in phase space
  becomes $N^{(2n-1)-2} > 1 (n\ge 2)$,
  where $N$ is the number of points sampled along a given axis.
  Thus, a naive estimation of the computational cost is
  on the order of ${(2^{N_\mathrm{res}})}^{(2n-1)-1}$
  if the cost of each slice is same as above: $2^{N_\mathrm{res}}$
  and $N=2^{N_\mathrm{res}}$.
  Not coincidentally, this exponent (=$(2n-1)-1$)
  is the same as the dimension of the reactivity boundaries, {\it i.e.,\/}
  $\mathcal{W}_\mathrm{asym}^\mathrm{(s)}$ and $\mathcal{W}_\mathrm{asym}^\mathrm{(u)}$.
  In Fig.~\ref{fig:cost} of Appendix~\ref{apx:Sampling},
  we illustrate how the order estimates surmised here
  correspond to the computational cost {seen} in practice.

  The output of the algorithm ---{\it i.e.}, the chain of straddling pairs---
  can be used to sample other parts of the boundary
  by propagating forward or backward in time (See Fig.~\ref{fig:steps} for example).
  However, due to the nature of a trajectory on a stable or unstable manifold,
  the distance between adjacent straddling pairs will exponentially increase by the propagation.
  Thus, depending on the integration time,
  we may need to obtain more pairs in between some pairs we already have.
  In such cases, we can use the output as an input to the first stage of the algorithm.
  We can then apply the bisection method $N_\mathrm{res}^\prime$ times father,
  and thereby obtain $2^{N_\mathrm{res}^\prime}$ times greater resolution than the input
  as the output.
  The interactive application of our algorithm
  thus results in an approximate curve representing
  the reactive boundary to a desired resolution
  (by the second stage) and
  precision (by the third stage.)

\section{NBC-ATI Analysis of Ketene Isomerization}\label{sec:Application}

\subsection{Reduced Ketene Model}\label{ssec:ketene}

  \begin{table}[t]
    \caption{%
      Parameters of the reduced ketene model of Eq.~\eqref{eq:potential} taken
      from Ref.~\onlinecite{gezelter1995}.
    }
    \begin{tabular}{lllll}
      \toprule
      Parameter & Value & Unit \\
      \midrule
      $a_2$ & $-2.3597\times 10^{-3}$ & $E_\mathrm{h}a_0^{-2}$\\
      $a_4$ & $ 1.0408\times 10^{-3}$ & $E_\mathrm{h}a_0^{-4}$ \\
      $a_6$ & $-7.5496\times 10^{-5}$ & $E_\mathrm{h}a_0^{-6}$ \\
      $c$ & $ 7.7569\times 10^{-3}$ & $E_\mathrm{h}a_0^{-2}$ \\
      $d$ & $1.9769$ & $a_0^{-2}$\\
      $m_\mathrm{F}$ & $9580.46$ & $m_\mathrm{e}$ \\
      $m_\mathrm{H}$ & $1837.1$ & $m_\mathrm{e}$ \\
      $k_1$ & $1.0074\times 10^{-2}$ & $E_\mathrm{h}a_0^{-2}$\\
      $d_1$ & $-2.45182\times 10^{-4}$ & $E_\mathrm{h}a_0^{-5}$ \\
      $k_2$ & $2.9044\times 10^{-2}$ & $E_\mathrm{h}a_0^{-2}$\\
      $d_2$ & $-8.5436\times 10^{-4}$ & $E_\mathrm{h}a_0^{-5}$ \\
      \bottomrule
    \end{tabular}\label{tbl:potential}
  \end{table}

  To illustrate the \ac{NBC-ATI} method,
  we apply it to the reaction dynamics of ketene
  to uncover its reaction geometry by \acp{ATI}.
  To this end,
  we use a reduced model of ketene introduced
  by Gezelter and Miller\cite{gezelter1995}
  and adopted by Craven and Hernandez to illustrate \acp{LD}.\cite{hern16d} 
  \begin{subequations}\label{eq:potential}
  \begin{flalign}
    V(q_\mathrm{F},q_1,q_2)
    :=V_\mathrm{F}(q_\mathrm{F})
    +\sum_{i=1,2} \frac{k_i}{2}{\left(q_i+d_i\frac{q_\mathrm{F}^4}{k_i}\right)}^2,\\
    V_\mathrm{F}(q_\mathrm{F})
    :=a_2 q_\mathrm{F}^2+a_4 q_\mathrm{F}^4+a_6 q_\mathrm{F}^6
    +c q_\mathrm{F}^2 \ee^{-d q_\mathrm{F}^2}.
  \end{flalign}
  \end{subequations}
  We use the parameters fitted by Gezelter and Muller\cite{gezelter1995}
  reproduced in Table~\ref{tbl:potential}
  with the correction on
  $m_\mathrm{F}$ by Ulusoy et al.\cite{hern14e}
  The fit is based
  on the result of an {\it ab initio\/} calculation at the level of
  CCSD(T)/6--311G({\it df,p\/}) by Scott et al.,\cite{schaefer94} %chktex 36
  where $q_\mathrm{F}$, $q_1$ and $q_2$ are coordinates of the systems.
  This model uses
  the normal mode coordinate of the oxirene geometry.\cite{gezelter1995}
  That is, $q_\mathrm{F}$ is the motion along
  the normal mode reaction coordinate,
  $q_1$ is the out-of-plane wagging and twisting motion
  of the hydrogens relative to CCO plane,
  and $q_2$ is the in-plane rocking and scissoring motion
  of the hydrogens relative to CCO plane.
  All the parameters are fixed to reproduce the structures of
  oxirene and formylmethylene intermediates.

  Ketene is known for its remarkable
  photo-isomerization and photo-dissociation behavior as first
  observed by Moore et al.\cite{moore91,moore92,moore93}
  Its unusual reaction dynamics was discussed in 
  the context of roaming reactions
  without\cite{hern13c,hern14e,Mauguiere2017}
  and with\cite{hern16d} the application of external forces.
  In the latter case, the interaction drives a dipole 
  along $q_\mathrm{F}$
  whose moment was obtained using
  B3LYP/6--311+G** by Craven and Hernandez.\cite{hern16d}
  The resulting potential interaction, and dipole moment can be
  written as
  \begin{subequations}\label{eq:dipole}
  \begin{flalign}
    V_\mathrm{ex}(q_\mathrm{F},t)=&\mathcal{E}_0 \sin(\omega t)\mu_m(q_\mathrm{F})\\
    \mu_m(q_\mathrm{F})
    :=&\mu_0
    (\ee^{-\alpha {(q_\mathrm{F}-q_0)}^4}+\ee^{-\alpha {(q_\mathrm{F}+q_0)}^4})\nonumber\\
    &+\mu_\mathrm{ketene}
  \end{flalign}
  \end{subequations}
  with the parameters reproduced in Table~\ref{tbl:dipole}.

  \begin{table}[t]
    \caption{%
      Parameters of the external force of Eq.~\eqref{eq:dipole}
      taken from Ref.~\onlinecite{hern16d}
    }
    \begin{tabular}{lllll}
      \toprule
      Parameter & Value & Unit \\
      \midrule
      $\mu_0$ & $0.546$ & $e a_0$\\
      $\mu_\mathrm{ketene}$ & $ 0.602$ & $e a_0$ \\
      $q_0$ & $1.95$ & $a_0$ \\
      $\alpha$ & $ 0.0701$ & $a_0^{-4}$ \\
      \midrule
      $\mathcal{E}_0$ & $0.03$ & $E_\mathrm{h} e^{-1} a_0^{-1}$ \\
      $\omega$ & $ 0.0025$ & $E_\mathrm{h} \hbar^{-1}$ \\
      \bottomrule
    \end{tabular}\label{tbl:dipole}
  \end{table}

  Below, we first demonstrate the \ac{NBC-ATI} method for the 1\ac{DoF}
  and 3\ac{DoF} ketene models under an external force.
  The Hamiltonian and \ac{EoM} for the 1\ac{DoF} system is:
  \begin{flalign}
    H_\mathrm{F}
    &=p_\mathrm{F}^2/(2m_\mathrm{F})
    +V_\mathrm{F}(q_\mathrm{F})+V_\mathrm{ex}(q_\mathrm{F},t), \\
    \frac{\dd}{\dd t}
    \left(
    \begin{array}{c}
      q_\mathrm{F} \\
      p_\mathrm{F}
    \end{array}
    \right)
    &=
    \left(%
    \begin{array}{r}
      \partial_{p_\mathrm{F}} H_\mathrm{F}\\
      -\partial_{q_\mathrm{F}}H_\mathrm{F}
    \end{array}
    \right),\label{eq:EoM1D}
  \end{flalign}
  where $p_\mathrm{F}$ is the conjugate momentum of $q_\mathrm{F}$, 
  and $m_\mathrm{F}$ is the particle mass at $q_\mathrm{F}$.
  For the 3\ac{DoF} system,
  the Hamiltonian and \ac{EoM} with a Langevin bath is:
  \begin{flalign}
    H
    &=\frac{p_\mathrm{F}^2}{2m_\mathrm{F}}
    +\sum_{i=1,2} \frac{p_i^2}{2m_\mathrm{H}}+V(q_\mathrm{F},q_1,q_2), \\
    \left(
    \begin{array}{c}
      \dd q_i \\
      \dd p_i
    \end{array}
    \right)
     &=
    \left(%
    \begin{array}{l}
      \partial_{p_i} H \dd t \\
      -\partial_{q_i}H \dd t-\gamma (p_i/m_i) \dd t+\dd B_{t}(t)
      \label{eq:EoM3D}
    \end{array}
    \right),%
  \end{flalign}
  where
  $i=1, 2, \mathrm{F}$,
  $p_i$ the conjugate momentum, $m_i$ the mass of $q_i$,
  $m_i=m_\mathrm{H}~(i=1,2)$,
  $\gamma=0.0025~E_\mathrm{h} \hbar^{-1}$ is the friction, and
  $\dd B_{t}(t)\sim \mathcal{N}(0,\kB T \gamma \dd t)$ 
  follows a Wiener process
  with the Boltzmann constant $\kB$ and the temperature $T=300~\mathrm{K}$.
  Here $\dd B_{t}(t)\sim \mathcal{N}(0,\kB T \gamma \dd t)$
  means that the random variable $B(t+\dd t)-B(t)$ follows
  normal distribution $\mathcal{N}(0,\kB T \gamma \dd t)$
  with average $0$ and variance $\kB T \gamma \dd t$.
  By definition,
  $\dd B_{t}(t)$ satisfies the stochastic-process version of
  the fluctuation-dissipation theorem.\cite{Oksendal2003}
  Equation~\eqref{eq:EoM3D} is deterministic for a stochastic instance.
  That is, Eq.~\eqref{eq:EoM3D} (or generally in It{\^ o} process)
  has pathwise uniqueness\cite{Oksendal2003} of the solution.
  For this reason,
  we use the same noise instance ${\{\dd B_{t}(t)\}}_t$
  for all the stochastic trajectories.

  Equation~\eqref{eq:EoM1D} is integrated numerically
  by the Dormand-Prince (at 5th order) method with step size control
  (for absolute and relative error is $10^{-13}$)
  implemented in the C++ \texttt{boost::numeric::odeint} library.\cite{Boost}
  Numerical integration of Eq.~\eqref{eq:EoM3D}
  is performed by the Euler-Maruyama (at 1st order) 
  method with $\dd t=0.01$.

\subsection{Asymptotic Trajectory Indicator}\label{ssec:ATI}

  \begin{figure}[t]
    \includegraphics[clip=true,width =\figurewide]{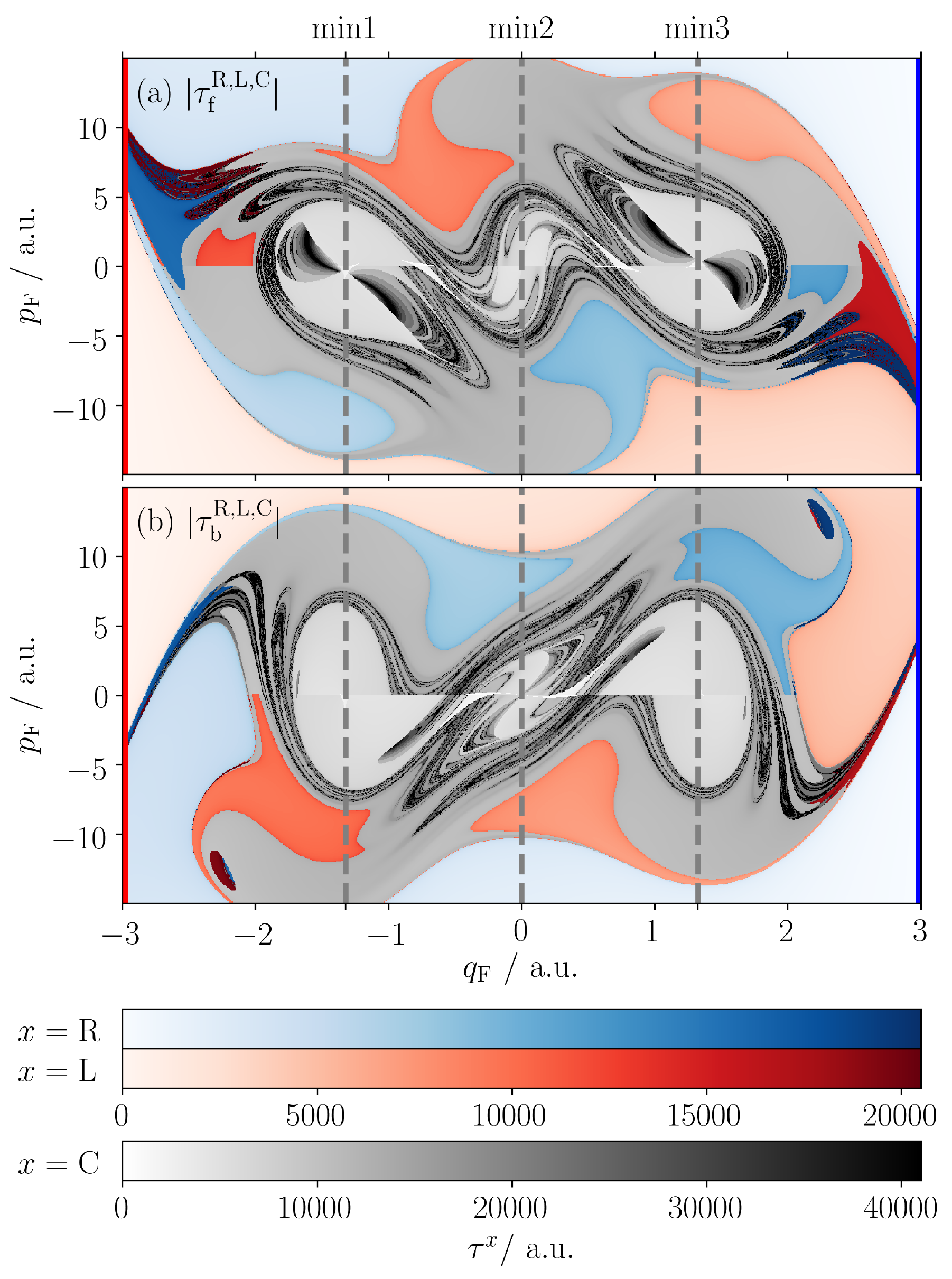}
    \caption{%
      Phase space representation of the ATI values for the 1D ketene model
      in an external field for the (a) forward $\tau_\mathrm{f}^\mathrm{R,L,C}$
      and (b) backward $\tau_\mathrm{b}^\mathrm{R,L,C}$ time propagation.
      The color for each initial condition
      is chosen according to the absorbed position:
      $q_\mathrm{F}=-3$ (red),
      $q_\mathrm{F}=3$ (blue),
      and
      the initial coordinate with the same velocity sign (gray),
      as also shown in Fig.\ref{fig:abs}.
    }\label{fig:global}
  \end{figure}

  We now demonstrate the usefulness of the \ac{ATI}
  for the 1 \ac{DoF} (Eq.~\eqref{eq:EoM1D})
  and 3 \ac{DoF} (Eq.~\eqref{eq:EoM3D}) ketene models.
  To this end,
  we use the visualization scheme explained
  in relation to
  Figs.~\ref{fig:HyperbolicFlow} and~\ref{fig:abs}.
  In Fig.~\ref{fig:global}, we show the result for the 1 \ac{DoF} model
  marking each location with the value of the 
  \ac{ATI} for trajectories starting at that point and ending when
  they reach $|q_\mathrm{F}|=3$.
  As can be seen, for the forward time propagation,
  trajectories starting from initial conditions
  at $q_\mathrm{F}=-3, p_\mathrm{F}<0$ ($q_\mathrm{F}=3, p_\mathrm{F}>0$)
  have the lightest red (blue) color,
  in Fig.~\ref{fig:global}a,
  because these trajectories are absorbed immediately.
  Similarly, in Fig.~\ref{fig:global}b,
  trajectories starting
  at $q_\mathrm{F}=-3, p_\mathrm{F}>0$ ($q_\mathrm{F}=3, p_\mathrm{F}<0$)
  have the lightest red (blue) color.
  The areas with large $|p_\mathrm{F}|$,
  not plotted in the figure, correspond to
  ballistic trajectories that
  will not be trapped in the wells.
  The areas which have darker reds  and blues
  correspond to trajectories
  that bounce back at the right and left external saddles
  $q_\mathrm{F}=2.8 (=-2.8)$, respectively,
  and then escape from it without a recurrence.
  These areas always present darker colors at the edge,
  due to the presence of asymptotic (long-time) trajectories.
  There is a line of discontinuities at $p_\mathrm{F}=0$ 
  due to the application of the recurrent condition, and 
  is an artifact of the way we define absorption.
  The gray colored area results from the limitation in the propagation time used in 
  our calculation to be insufficiently long to resolve
  these areas in terms of red and blue absorbing boundaries.
  For example, in the case when we apply the recurrent condition for the later time starting at $\omega t=14 \pi$,
  some gray areas in Fig.~\ref{fig:global}a
  are colored as it is seen in Fig.~\ref{fig:highlight}b.
  We show below that the manifold-like structures in this area
  are in fact a manifold.

  \begin{figure}[t]
    \includegraphics[clip=true,width =\figurewide]{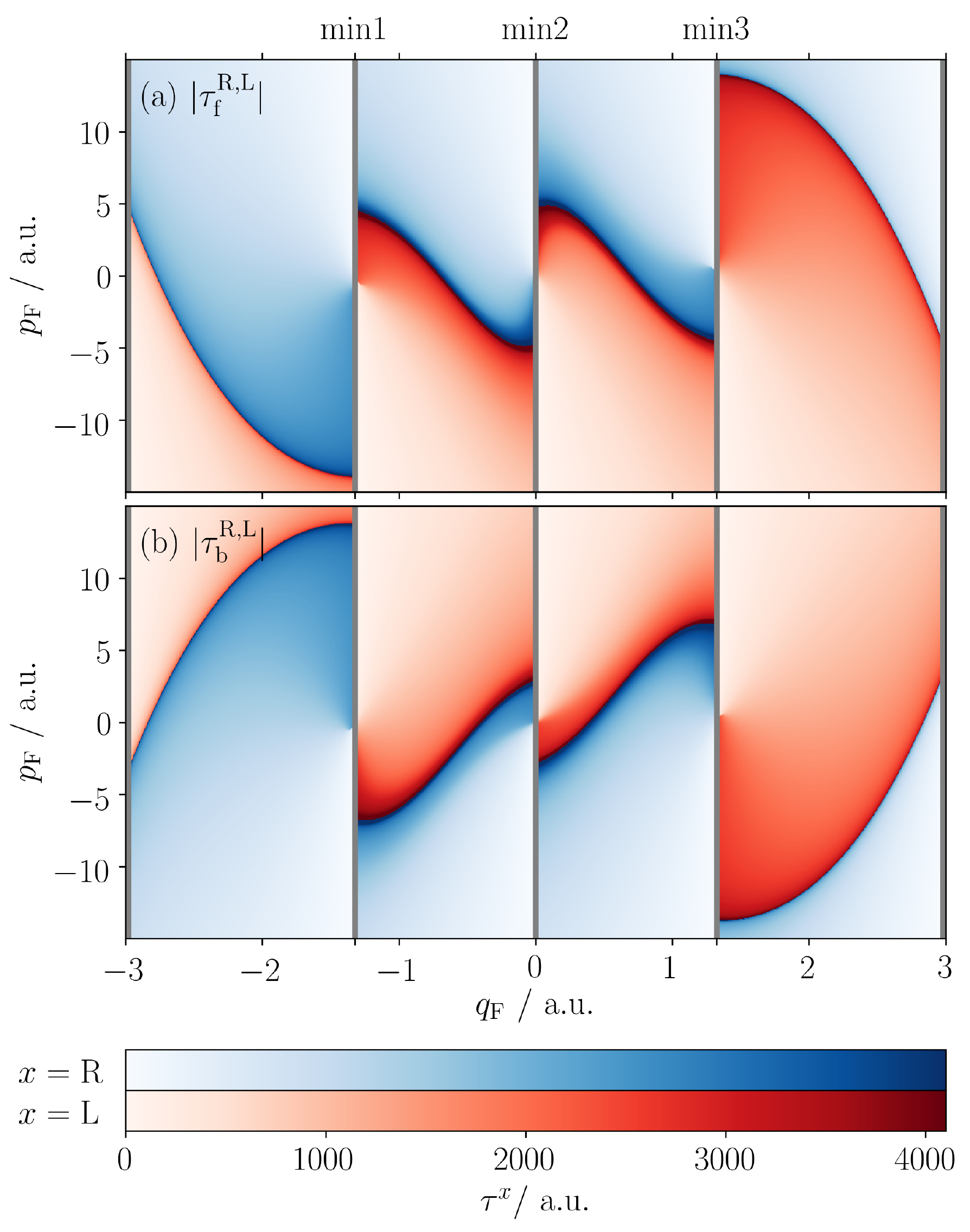}
    \caption{%
      Phase space representation of the ATI values for the 1D ketene model
      in an external field for the (a) forward $\tau_\mathrm{f}^\mathrm{R,L}$
      and (b) backward $\tau_\mathrm{b}^\mathrm{R,L}$ time propagation.
      The color for each initial condition
      is chosen according to the absorbed position:
      left (red),
      and
      right (blue).
      The absorbing boundaries are located at
      $q_\mathrm{F}=-3$, the local minima (min1, min2, min3),
      and $q_\mathrm{F}= 3$ are shown in gray
      corresponding to the red or blue boundaries of Fig.~\ref{fig:abs}
      within a given region.
    }\label{fig:local}
  \end{figure}

  As discussed in Sec.~\ref{sec:Theory},
  the use of the \ac{ATI} to locate the stable or unstable 
  manifold of a \ac{NHIM} in a given region
  becomes unclear when the region contains more than one \ac{NHIM}
  because of the challenge in assigning asymptotic behavior
  to a particular origin.
  To address this challenge,
  we consider the effect of inserting additional
  absorbing boundaries.
  For example, the results 
  corresponding to the 
  insertion of absorbing boundaries at
  the potential energy minima (min1, min2, min3 
  defined in the caption to Fig.~\ref{fig:highlight}
  are shown in Fig.~\ref{fig:local}.
  The identification of the manifold results from propagation
  of trajectories that are 5 times smaller than the previous figure (as indicated by the
  smaller \ac{ATI} values).
  The increased efficiency results from the fact that
  we do not need to use the recurrent condition
  due to the presence of the additional absorbing conditions.
  We are thus able to observe the manifolds that mediate internal mechanism
  in addition to the escaping process.
  Below in Subsec.~\ref{ssec:Turnstile} and Appendix~\ref{apx:Sampling},
  we demonstrate the resolution of the entire phase space
  using this approach.

  \begin{figure}[t]
    \includegraphics[clip=true,width =\figurewide]{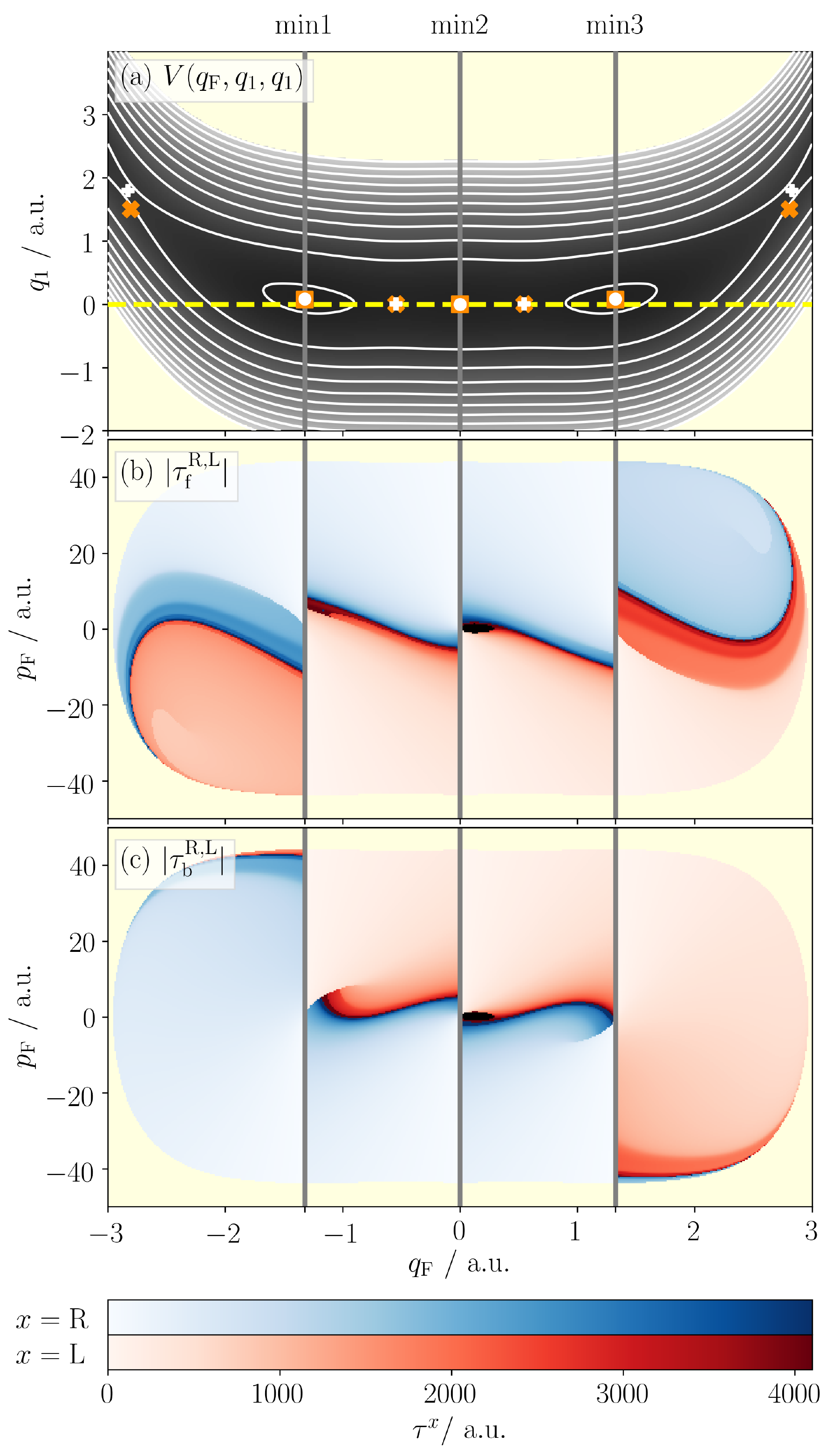}
    \caption{%
      (a)
      Potential energy surface for the reduced 3D ketene model on the $q_2=q_1$ plane.
      The white filled-circle symbols (plus symbols)
      indicate the position of minima (saddles) on the plane,
      and the orange filled-square symbols (times symbols)
      correspond to projected minima (saddles).
      The yellow dashed line is the surface of initial conditions.
      The middle and the bottom panels are
      the forward (b), and backward (c)
      \acp{ATI} on the $q_1=q_2=0$ plane.
      The absorbing boundaries are the same as in Fig.~\ref{fig:local}
      except for the outer most boundaries that are now located at $q_\mathrm{F}=\pm 4$.
      The light yellow areas in (b) and (c) are
      on the outside of the initial energy $E=0.1~\mathrm{a.u.}$.
    }\label{fig:local3D}
  \end{figure}

\begin{figure}[t]
    \includegraphics[clip=true,width =\figurewide]{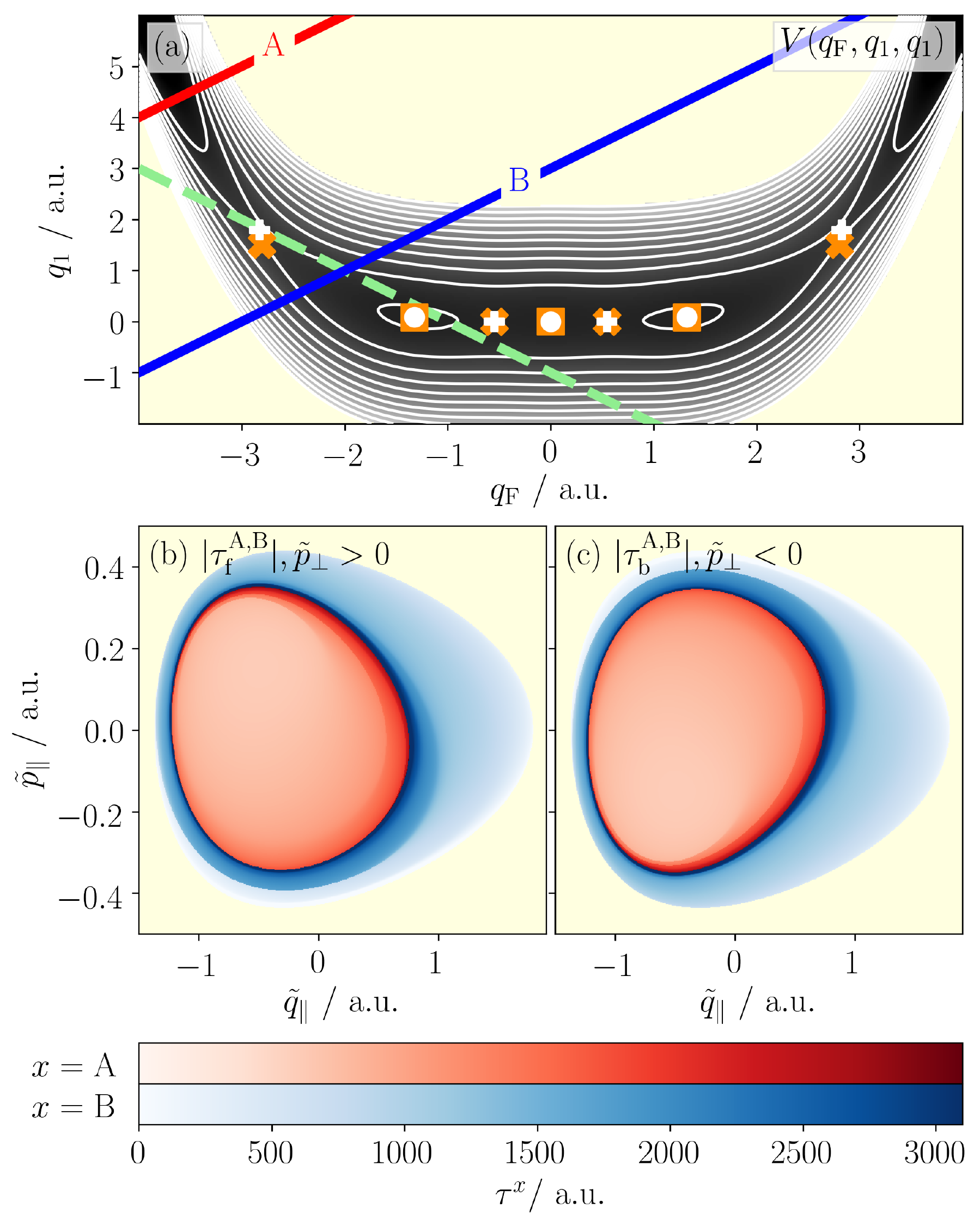}
    \caption{%
      (a)
      The potential energy surface of reduced 3D ketene model on $q_1=q_2$ plane
      (see Fig.~\ref{fig:local3D}) and superimposed absorbing boundaries
      $A$ (red line) and $B$ (blue line).
      The green dashed line corresponds to $\tilde{q}_1=0$.
      (b) forward and (c) backward \acp{ATI}
      of reduced 3D ketene model
      in a Langevin bath
      with initial conditions on
      surface $B$ with $q_1=q_2$.
      The light yellow areas in (b) and (c) are
      the outside of the initial energy $E=0.1~\mathrm{a.u.}$.
    }\label{fig:ext3D}
  \end{figure}

  The \ac{ATI} can be used in systems
  that are stochastic/non-autonomous, with multiple \acp{DoF},
  and without 
  need for an {\it a priori\/} reaction coordinate.
  To demonstrate this fact,
  we present a visualization of \ac{ATI} for the 3 \acp{DoF} ketene model 
  coupled to a Langevin bath.
  In Fig.~\ref{fig:local3D}a and  Fig.~\ref{fig:ext3D}a,
  the $q_1=q_2$ slice contour plot of the potential energy surface is shown.
  The white filled-circles (plus symbols) are local minima (saddles) on the slice,
  and the orange filled-squares (times symbols) are projected local minima (saddles).
  In Fig.~\ref{fig:local3D},
  the initial conditions are prepared on the yellow dashed line
  with positive out-of-slice velocity and constant energy
  $E=0.1~\mathrm{a.u.}$ ($q_1=q_2=0$ and $p_1=p_2>0$).
  The absorbing conditions are the same as in Fig.~\ref{fig:local}
  except the outermost boundaries $q_\mathrm{F}=\pm 3$
  that are now $q_\mathrm{F}=\pm 4$ and which are not shown in the figure.
  This change in the outer boundaries was necessitated
  % I think necessitated is better, because the change is consequence of the observation
  by the observed discontinuities in the \ac{ATI}
  as implemented with the narrower
  absorbing boundaries at $q_\mathrm{F}=\pm 3$,
  possibly due to the overlap of the boundary with the \acp{NHIM}.
  Since we use the fixed initial energy,
  there is an upper and lower limit for the velocity not seen 
  in Fig.~\ref{fig:local}.
  In comparison with
  Fig.~\ref{fig:local},
  there is no change in the timescale of the \acp{ATI} in Fig.~\ref{fig:local3D}
  because the increase of the \acp{DoF} does not affect the timescale of the motion.

  Although the reaction coordinate on the potential energy function is curved,
  we are still able to locate the reactivity boundaries on each cell
  According to the result of the \ac{ATI},
  the initial conditions are best given by $p_1>0$ ($p_1<0$)
  for positive (negative) time integration.
  This fact can be observed from the backward time propagation case,
  shown in the left cell of Fig.~\ref{fig:local3D}c.
  Therein, we mostly observe bounce back trajectories
  from the right (min1) shown by blue colors.
  This occurs since we take $p_1>0$.
  As a consequence, trajectories have opposite velocity along $p_1$
  in comparison with the 
  trajectories sliding down from the left external saddle.
  The right cell of  Fig.~\ref{fig:local3D}c has similar behavior
  due to the symmetry of the potential energy surface.

  The black areas in the middle right cells of 
  Figs.~\ref{fig:local3D}b and~\ref{fig:local3D}c 
  correspond to trajectories
  that have not finished in the computation time $t=10^5$.
  Those trajectories are, in fact, those
  which stay vibrating on the initial coordinate $q_\mathrm{F}$ plane,
  possibly in the attraction basin of the fixed points/invariant manifold.
  There is a line in Fig.~\ref{fig:local3D}c (center cells)
  of small discontinuities
  which is an artifact of the absorbing boundaries.
  The reactivity boundary must have a singular value of $\tau$,
  otherwise it appears due to an inappropriate absorbing boundary.

  Beside the absorbing boundaries considered in Fig.~\ref{fig:local3D},
  for the multiple \acp{DoF} systems, one can use a variety of absorbing boundaries,
  such as a boundary transverse to the reaction path
  or the one used in the reaction island theory.\cite{deLeon1}
  In Fig.~\ref{fig:ext3D}, we present the result for the absorbing boundaries
  A (red) and B (blue) that enclose the left external saddle point,
  and are given by $q_1=q_\mathrm{F}+8$ and $q_1=q_\mathrm{F}+3$ respectively.
  The initial conditions are prepared on the boundary B with $q_1=q_2$
  by using the mass-weighted coordinates
  along B ($\tilde{q}_{\parallel}$)
  and orthogonal to B ($\tilde{q}_{\perp}$).
  We define the zero axis ($\tilde{q}_1=0$)
  by $q_1=-q_\mathrm{F}-1$ (green dashed line),
  which is along the coordinate $\tilde{q}_{\perp}$.
  Thus, the origin $\tilde{q}_1=0$ of Fig.~\ref{fig:ext3D} is
  at $(q_\mathrm{F},q_1,q_2)=(-2,1,1)$,
  which is the crossing point of the line B and the dashed green line.
  Here we use a Lagrangian transformation
  to obtain $\tilde{q}_\parallel$ and $\tilde{q}_\perp$
  under the condition $p_1=p_2$.
  The result for the ATI is shown in Figs.~\ref{fig:ext3D}b and~\ref{fig:ext3D}c.
  In these panels, the reactivity boundary between blue and red area can be located
  even in the challenging case presented by a Langevin bath.
  Initial conditions with $\tilde{q}_1<0$ and $\tilde{p}_1<0$ ($\tilde{p}_1>0$)
  corresponding mostly to reacting trajectories (red) in 
   the forward or backward time propagation
  as seen in Fig.~\ref{fig:ext3D}b (Fig.~\ref{fig:ext3D}c).
  This is because these initial conditions are closer to the left external saddle
  and have velocities ahead to or from this saddle, respectively.

\subsection{Turnstile and Reaction Path}\label{ssec:Turnstile}

  \begin{figure*}[t]
    \includegraphics[clip=true,width =\figurewide]{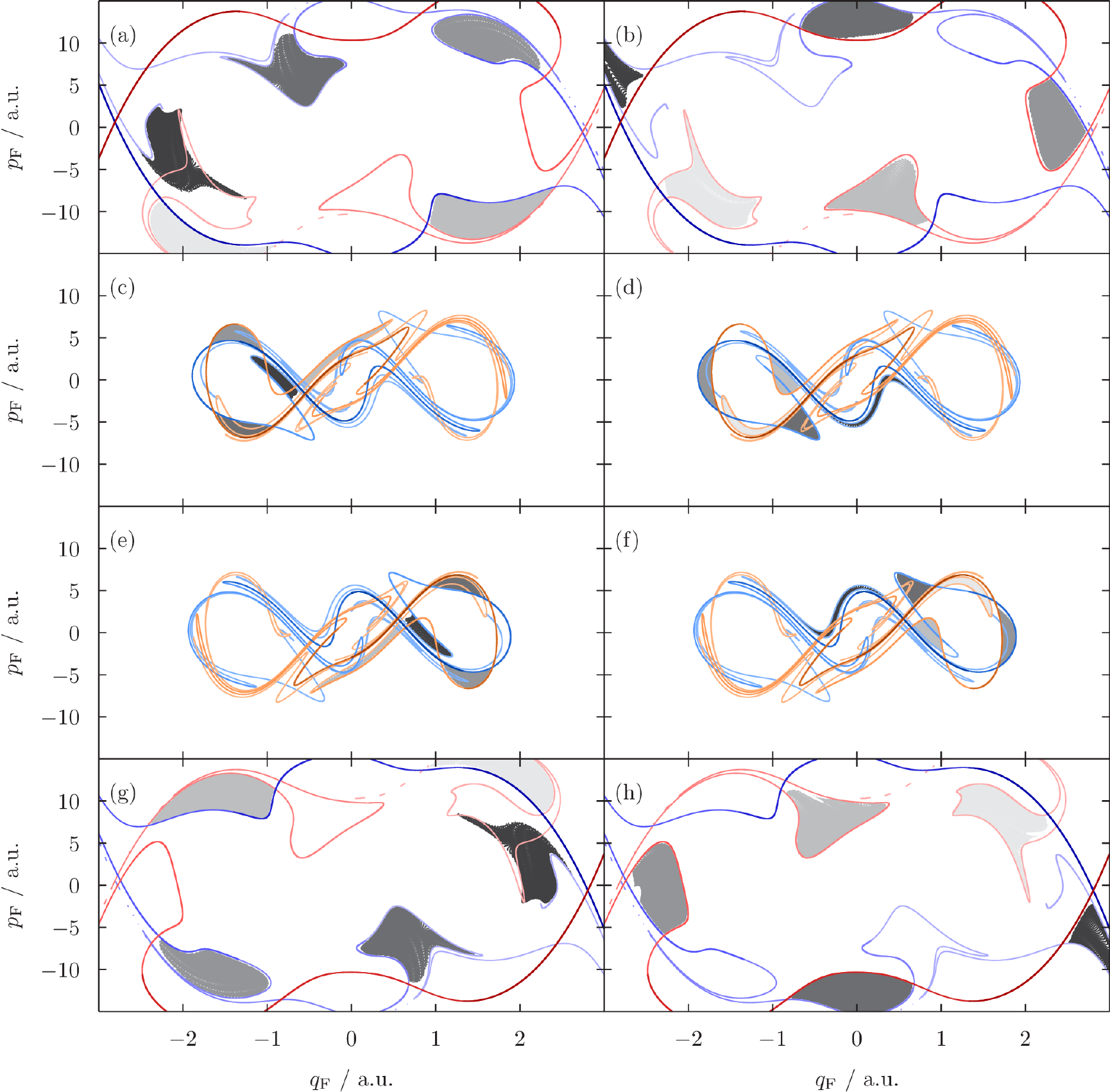}
    \caption{%
      The turnstile (reaction pathway) of each NHIM (saddle),
      mediated by the manifolds obtained 
      at phase $\omega t\equiv 0 \pmod{2\pi}$ of the 1D ketene model driven by an external force.
      The stable (blue, cyan) and unstable (red, orange) manifolds
      are drawn up to $|\omega t|= 8\pi$.
      The manifolds
      associated with the saddles
      for left external (a), right external (b), 
      left internal (g) and right internal (h)
      are shown in (c), (d), (e), (f) respectively.
      The colors of the manifolds become lighter, as $|t|$ increases.
      Left and right panels correspond to exiting and entering 
      trajectories, respectively, 
      Coherent sets of these trajectories are marked in gray,
      with increasingly lighter shades indicating later phases, 
      that are enclosed by
      the stable and unstable manifold.
    }\label{fig:turnstiles}
  \end{figure*}

  \begin{table}[t]
    \caption{%
      Four types of phase space regions separated by stable and unstable manifolds
    }
    \begin{tabular}{|l|l|l|l|} %chktex 44
      \cline{3-4}
      \multicolumn{2}{c|}{} & \multicolumn{2}{c|}{stable manifold} \\
      \cline{3-4}
      \multicolumn{2}{c|}{} & inside & outside \\
      \hline %chktex 44
      unstable & inside & (1) staying inside & (2) entering into\\
      \cline{2-4}
      manifold & outside & (3) exiting from & (4) staying outside \\
      \hline %chktex 44
    \end{tabular}\label{tbl:turnstile}
  \end{table}

  We now demonstrate a way to use
  the manifolds obtained by \ac{NBC-ATI} in Sec.~\ref{ssec:Algorithm}
  in the context of 
  turnstile\cite{Mackay1984} or lobe dynamics.\cite{Wiggins1990}
  As discussed in Fig.~\ref{fig:HyperbolicFlow},
  the stable and unstable manifolds is
  the destination and origin dividing boundaries
  respectively, of the dynamics.
  The areas divided by the reactivity boundaries
  are categorized into four types (see Table~\ref{tbl:turnstile}):
  trajectories that are
  (1) staying inside, (2) entering into,
  (3) exiting from, and (4) staying outside the trapping area (a chemical state).
  Among the four types, an area which encloses the reaction pathway
  is categorized into (2) or (3)
  and corresponds to a time slice of the pathway.

  In Fig.~\ref{fig:turnstiles},
  the manifolds produced from each cell in Fig.~\ref{fig:local}
  are shown through the superposition of the manifolds at
  the phases $|\omega t|= 2\pi M (M=0,\dots,4)$
  (see Appendix~\ref{apx:Sampling}).
  These manifolds correspond to the stable (blue, cyan)
  and the unstable (red, orange) manifolds of the \ac{NHIM} in each cell.
  To expose the reaction mediated by these manifolds,
  we draw a set of trajectories prepared in an enclosed area (gray).
  These trajectories move coherently
 %   I use ``coherent'' and ``coherently'' in the same terms of Lagrangian Coherent Structures
  from one enclosed area to another
  and are shown (stroboscopically) at 
  $\Delta t=2\pi/\omega$ time intervals with changing strength of its color.
  Darker color indicate points captured in earlier time periods.
  As the color becomes lighter,
  the set of trajectories goes out of (or into) the trapped area
  in the left (or right) column in Fig.~\ref{fig:turnstiles}.
  Although, we illustrate trajectories in a few selected area,
  this should suffice to observe that
  the dynamics are mediated by the manifolds as expected.

  \begin{figure*}[t]
    \includegraphics[clip=true,width =\figurewide]{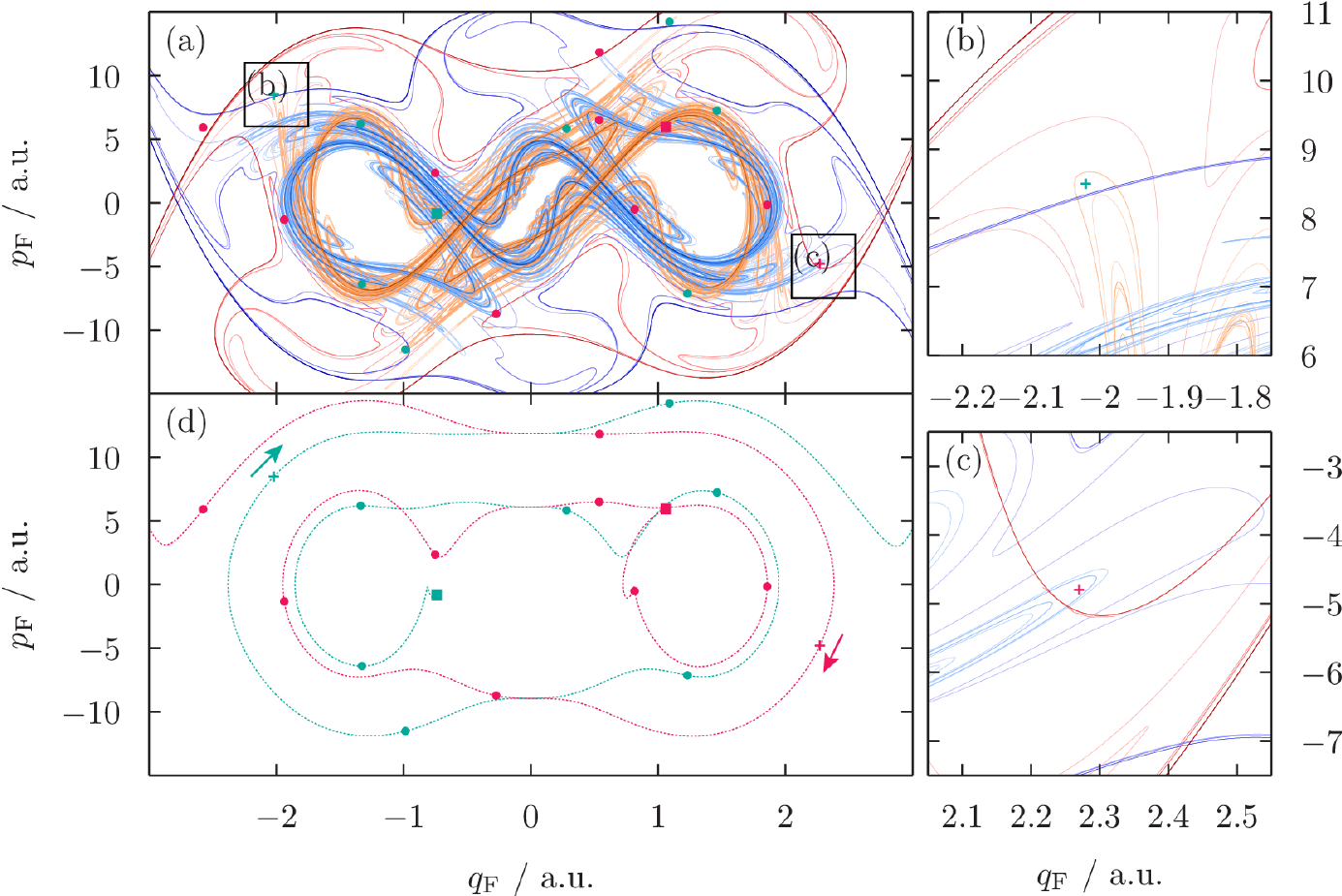}
    \caption{%
      The intersections of the internal and external stable/unstable manifolds
      of ketene 1 D with external field.
      (a) Superposition of manifolds at phase $\omega t \equiv 0 \pmod{2\pi}$
      up to $|\omega t|= 14\pi$
      which partially shown in Fig.~\ref{fig:turnstiles}.
      (b) and (c) Magnifications of areas marked in (a)
      where we sample the initial conditions of trajectories (plus symbols).
      (d) The sampled entering (pink dotted line) and exiting (green dotted line) trajectories.
      The plus symbols, filled-squares, and filled-circles are
      time-slices at the phase $\omega t \equiv 0~\pmod{2\pi}$
      of trajectories (dotted lines).
    }\label{fig:getin}
  \end{figure*}

  The manifolds in the left and right column
  of Fig.~\ref{fig:turnstiles} are the same
  but the initial gray-colored areas are different
  as they correspond to the time slice of
  the exiting and entering reaction paths, respectively.
  They illustrate the last or first few steps of the reaction pathways.
  One can combine these pictures to see a reaction pathway
  exiting into (entering from) the internal well
  from (to) the outside of the observed area.
  Such a pathway must be explained by the intersection of
  the reaction pathways mediated by
  internal ((b) and (c)) and external ((a) and (d)) manifolds in Fig.~\ref{fig:turnstiles}.
  To see the intersection,
  we draw the manifolds up to the phase $M=7$ in Fig.~\ref{fig:getin}.
  We uncover two initial conditions that are enclosed
  by the both internal and external manifolds
  in Fig.~\ref{fig:getin}b and c respectively.
  The resulting pink (green) trajectory shows that pathway
  comes into (goes out)
  the internal well directly from (toward) outside.
  These correspond to a case in which
  there is a fast cooling-down (fast excitation)
  process due to the presence of an external field.
  However,
  from the size of the enclosing area of the initial conditions
  in Fig.~\ref{fig:getin}b and c,
  one can also observe that the
  amount of such coherent initial conditions are small
  indicating that the event is pretty rare.
  Thus, there is a time scale separation for transitions between
  internal trapping and external trapping trajectories
  since the former has less energy than the latter.

  \begin{figure}[t]
    \includegraphics[clip=true,width =\figurewide]{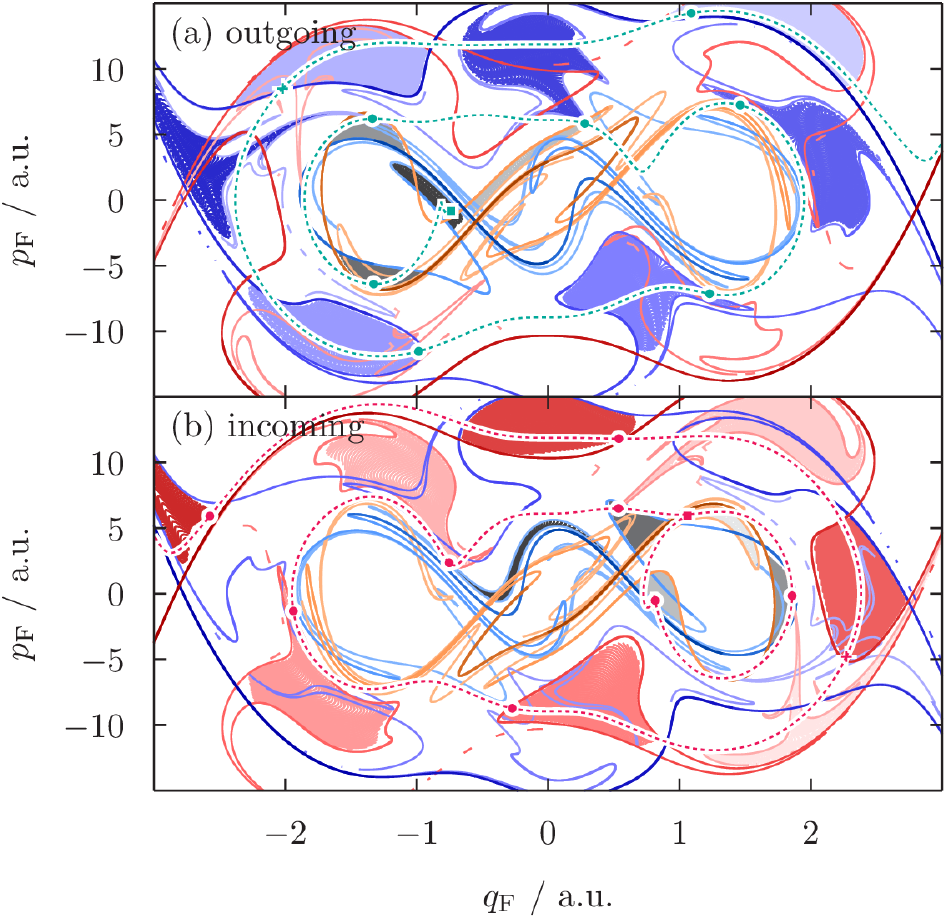}
    \caption{%
      (a) The exiting trajectories of 1D ketene
      shown in Fig.~\ref{fig:getin}(d) %chktex 36
      superimposed on the associated coherent set of trajectories
      shown in Fig.~\ref{fig:turnstiles}
      (c) (gray) and (g) (blue).
      (b) the same for the entering trajectory
      superimposed on the associated coherent set of trajectories
      shown in Fig.~\ref{fig:turnstiles}
      (b) (red) and (f) (red).
      The color of the set of the trajectories get lighter
      when the time period is later.
    }\label{fig:combine}
  \end{figure}

  In Fig.~\ref{fig:combine},
  we interpret the trajectories of Fig~\ref{fig:getin} 
  in the context of turnstiles.
  In Fig.~\ref{fig:combine}a, the limit point
  ---shown as a green filled-square---
  is in the area colored by the darkest gray.
  As time progresses,
  the green filled-circles move into ever lighter colored areas in the Poincar{\'e} map.
  In the first three periods,
  the circles from the green filled-square
  all moved into areas shaded with increasingly lighter gray.
  Starting with the second period, the circles also move into blue areas.
  These later circles move into areas with increasingly lighter blue shades
  until they move outside of the figure.
  Similarly, the pink trajectory starting from a different point in
  the phase space experiences a series of 
  gray and red areas in Fig.~\ref{fig:combine}b upon application of the
  Poincar{\'e} map.
  Therefore, the trajectories sampled at the points in Fig.~\ref{fig:getin} are
  successfully identified by the intersection of the reaction pathways mediated by
  internal ((b) and (c)) and external ((a) and (d)) 
  manifolds in Fig.~\ref{fig:turnstiles}.

  Finally, let us revisit Fig.~\ref{fig:highlight}.
  The \ac{ATI} shown in 
  Fig.~\ref{fig:highlight}b 
  was computed 
  without applying the absorption of recurrent trajectories
  until $|\omega t|>7\times 2\pi$.
  The choice for this max time 
  is in agreement with the number $(=7)$ of periodic lobes
  (red or blue colored areas) of the
  coherent trajectories that we followed
  in Fig.~\ref{fig:combine}.
  The coherent structures, mediated by the reactivity boundaries
  and revealed by the \ac{NBC-ATI} method,
  are clearly visible
  in Fig.~\ref{fig:highlight}b
  because we compute ATI values for longer times
  than those shown Fig.~\ref{fig:global}.
  The yellow lines are the superpositions of
  the external stable manifolds
  that are also shown in Fig.~\ref{fig:getin}a as blue lines.
  These coherent structures agree with the 
  those shown in Fig.~\ref{fig:highlight}c 
  obtained directly through the use of global manifolds.
  That is, the phase space skeletons are correctly extracted 
  by the \ac{NBC-ATI} algorithm.
  It leads to consistent and correct implications on the dynamics.
  Therefore, all the information about reactivity is
  extracted only 
  from the manifolds obtained by the \ac{NBC-ATI} algorithm.

\section{Discussion}\label{sec:Discussion}

  Here, we analyze the possible use of the \ac{NBC-ATI} method to describe
  more general chemical reactions based on our findings from the analysis
  of the 1D and 3D ketene models under various coupling conditions.
  In Fig.~\ref{fig:local3D} and Fig.~\ref{fig:ext3D},
  we showed a 2D slice of the \ac{ATI} in the 3 \acp{DoF} phase space.
  To obtain them in the full phase space,
  naively one needs to achieve it for all the remaining slices.
  However, as we showed in Sec.~\ref{ssec:Turnstile},
  one can use a periodic identity of the dynamics
  to obtain samples by integrating
  across several period(s) of time. %chktex 36
  This sampling is more efficient for autonomous systems
  because the phase space is identical for all time.
  Although this type of identity reduces computational costs,
  the minimum computation costs must be proportional 
  to the dimension of the manifold
  in any numerical analysis.
  This is because
  even if we just uniformly sample a known $n$-dimensional manifold,
  a number of points
  proportional to the order of the power $n$ is needed.
  This is a fundamental limitation of numerical sampling techniques
  that perturbation theories do not suffer.

  Unlike in autonomous or periodic dynamical systems,
  there exists no natural Poincar{\'e} map
  in systems driven by aperiodic or stochastic differential equations.
  Hence, to see the reactivity boundaries at another time,
  one does not have any resource beyond the time-propagated manifold.
  In addition, for stochastic differential equations,
  the computational cost is larger
  because the integrators available for 
  such systems are not as efficient
  as those for an \ac{ODE}.
  To achieve a given resolution in the time-propagated manifold,
  the resolutions used within steps of the \ac{NBC-ATI} must 
  be chosen carefully ensuring that the computation is efficient.
  Due to the nature of a trajectory on a stable or unstable manifold,
  the distance between adjacent straddling pairs will
  exponentially increase by the time backward or forward propagation, respectively.
  Nevertheless, 
  the weighted sampling\cite{Nagahata2013a} is known to improve the efficiency.
  Since the \ac{NHIM} and their stable and unstable manifolds are smooth,
  interpolation between the straddling pairs will reduce computational cost 
  to some extent.
  For engineering purposes, this types of solutions can improve efficiency,
  although, what types of interpolation are allowed to use is still in question.
  Although the \ac{NHIM} and its stable and unstable manifolds are of great importance,
  there is no guarantee that
  that reactivity is always mediated by them.
  It appears in the study of a reaction associated
  with higher index saddles\cite{Nagahata2013a,komatsuzaki13a}
  that the manifold,
  which is not orthogonal to the most repulsive nor attractive direction,
  can produce reactivity boundaries.
  However, one should be careful
  to correctly ascribe the physical interpretation of these reactivity boundaries.
  They may not persist under perturbation.
  That is, a small perturbation {\it e.g.}, a small difference of potential energy surface,
  may introduce a large dynamical difference.
  The \ac{NBC-ATI} also provides a useful reference for determining
  which terms in perturbation theories should be retained.
  For example, normal form theories are known to be asymptotic series
  which necessarily diverge if one includes all terms in the expansion.

\section{Conclusion}\label{sec:Conclusion}

  In this paper,
  we have presented a formulation for the \ac{ATI}
  and an identification of reactivity boundaries\cite{Nagahata2013a}
  based on dynamical systems theory
  by revisiting the reactivity map.\cite{Wall1958,Wall1961,Wall1963,
  Wright1975,Wright1976,Wright1977,Wright1978,Laidler1977,Tan1977}
  To this end, we developed the \ac{NBC-ATI} method
  which effectively requires computational resources that are proportional
  to the dimensionality of the manifolds.
  We demonstrate the feasibility and efficiency of this
  approach on a reduced-dimensional ketene model\cite{gezelter1995}
  in 1D with external field, and in 3D coupled to a Langevin bath.
  The \ac{NBC-ATI}  method can address
  irregular reactions
  which are not accessible
  to conventional perturbation theories or other types of numerical analysis.
  Examples include
  the existence of roaming pathways,\cite{bowman2011,Mauguiere2017}
  bifurcation of the periodic orbit dividing surface,\cite{pollak78,komatsuzaki06a}
  dynamical switching of the reaction coordinate,\cite{komatsuzaki11}
  and a reaction associated with 
  higher index saddle(s).\cite{Nagahata2013a,komatsuzaki13a} %chktex 36

  In the reduced 1D ketene model,
  we obtained complex reaction paths
  based on turnstiles.\cite{Mackay1984}
  The turnstiles consist of stable and unstable manifolds
  associated with the \acp{NHIM} around the four potential energy saddle points.
  We find the internal trapping area in a chaotic sea
  corresponding to the two formylmethylenes and oxirene conformations.
  Using these phase space structures,
  we identified and characterized rare trajectories
  which are trapped by and escape from the internal wells in a short time.
  We thus demonstrated that our method has sufficient accuracy
  to reproduce the conventional analysis when it is accessible,
  and generalizes to reactions with more complex reaction geometry 
  as listed above.

  We have shown the applicability of the \ac{NBC-ATI} method
  for stochastic and higher dimensional systems through an application
  to a reduced 3D ketene model.
  We found that
  if one can identify an area
  which includes the \ac{NHIM} or an asymptotic manifold $\mathcal{M}_\mathrm{asym}$
  as a seed of the reactivity boundaries,\cite{Nagahata2013a}
  then the \ac{ATI} still allows us to identify the reactivity boundaries
  in stochastic and higher dimensional systems.
  Besides the advantage of reduced computational costs,
  the \ac{ATI} allows us to identify structures associated with the reaction path
  in cases 
  which are beyond reach to
  the conventional perturbation theories and numerical analysis.

\section*{Acknowledgments}\label{sec:Acknowledgments}
This work was partially supported by the National Science
Foundation (NSF) through Grant No.~CHE-1700749.
This collaboration has also benefited from support by the European
Union's Horizon 2020 Research and Innovation Program under the Marie
Sklodowska-Curie Grant Agreement No.~734557.

\appendix
\renewcommand\thefigure{\thesection.\arabic{figure}}
\setcounter{figure}{0}

\section{\ac{NHIM} persistence theorem}\label{apx:NHIMtheorem}

  The \ac{NHIM} is a multidimensional generalization of a hyperbolic fixed point
  such as that associated with the saddle point 
  ---{\it viz\/} the naive \ac{TS}--- in a chemical
  reaction.
  Here, we recapitulate the statement of the 
  theorem for the persistence of the 
  \ac{NHIM} under perturbations as
  originally proven by Fenichel,\cite{Fenichel72} and generalized by others.\cite{Hirsch1977,Eldering2013}
  For simplicity, we call this the \ac{NHIM} persistence theorem.

  \begin{figure}[t]
    \includegraphics[clip=true,width =\figurewide]{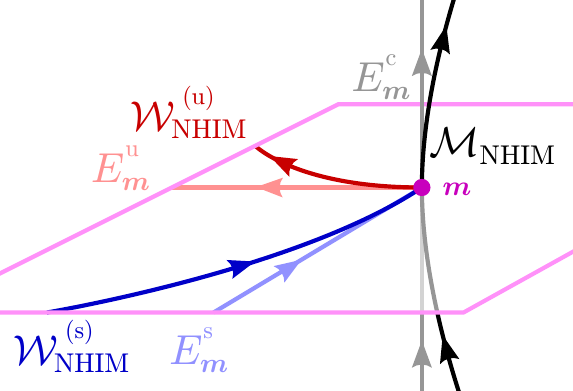}
    \caption{%
      Invariant splitting at $\bm{m}\in\mathcal{M}_\mathrm{NHIM}$:
      $E^\mathrm{c}_{\bm{m}}
      \otimes E^\mathrm{s}_{\bm{m}}\otimes E^\mathrm{u}_{\bm{m}}$
      and its relation with $\mathcal{M}_\mathrm{NHIM}$,
      $\mathcal{W}_\mathrm{NHIM}^\mathrm{(s)}$, and
      $\mathcal{W}_\mathrm{NHIM}^\mathrm{(u)}$.
    }\label{fig:NHIM}
  \end{figure}
  Suppose that in a given system, we have identified
  a smooth Riemannian manifold $\mathcal{Q}$,
  a flow on $\mathcal{Q}$: $\bm{\phi}^t\in C^k\ (r\ge 1)$,
  and a compact submanifold of $\mathcal{Q}$: $\mathcal{M}_\mathrm{NHIM}$.
  This manifold, $\mathcal{M}_\mathrm{NHIM}$, is \iac{NHIM} when:
  \begin{enumerate}
    \item $\mathcal{M}_\mathrm{NHIM}$ is invariant,
    {\it i.e.}, $\bm{\phi}^t(\mathcal{M}_\mathrm{NHIM})=\mathcal{M}_\mathrm{NHIM}$,
    \item There exists continuous splitting
      for $\forall \bm{m}\in\mathcal{M}_\mathrm{NHIM}$,
    \begin{equation}
      T_{\bm{m}} \mathcal{Q}=
      E^\mathrm{c}_{\bm{m}}
      \otimes E^\mathrm{s}_{\bm{m}}
      \otimes E^\mathrm{u}_{\bm{m}},
    \end{equation}
    of the tangent bundle of $\mathcal{Q}$ at $\bm{m}$
    with globally bounded, continuous associated projections:
    $\bm{\pi}_{\bm{m}}^\mathrm{c}$, $\bm{\pi}_{\bm{m}}^\mathrm{s}$,
    and $\bm{\pi}_{\bm{m}}^\mathrm{u}$
    such that splitting is invariant under the linearized flow
    \begin{equation}
      D_{\bm{m}}\bm{\phi}^t(E^x_{\bm{m}})=
      E^x_{D_{\bm{m}}\bm{\phi}^t(\bm{m})}
      \label{eq:splitting}
    \end{equation}
    for $\forall \bm{x}\in E^x~(x=\mathrm{c, s, u})$
    where $E^\mathrm{c}_{\bm{m}}=T_{\bm{m}}\mathcal{M}_\mathrm{NHIM}$ and
    $D_{\bm{m}}\bm{\phi}^t$ is differential of $\bm{\phi}^t$ at $\bm{m}$,
    {\it e.g.}\ the flow determined by the normal mode for the Hamiltonian systems.
    \item There exists constants
    $-\lambda_{\bm{m},\mathrm{s}}<-\lambda_{\bm{m},\mathrm{c}}\le 0
    \le \lambda_{\bm{m},\mathrm{c}}<\lambda_{\bm{m},\mathrm{u}}$,
    $c_{\bm{m},\mathrm{c}}, c_{\bm{m},\mathrm{s}}, c_{\bm{m},\mathrm{u}}\ge 1$,
    such that
      for the matrix
      $D_{\bm{m}}\bm{\phi}^t={(\partial_{x_i}\phi^t_j(\bm{x})|_{\bm{x}=\bm{m}})}_{ij}$
    \begin{eqnarray}
      &\forall t,\ \bm{x}\in E_{\bm{m}}^\mathrm{c}:&\nonumber\\
      &\norm{D_{\bm{m}}\bm{\phi}^t \bm{\pi}_{\bm{m}}^\mathrm{c}(\bm{x})}
      &\leq c_{\bm{m},\mathrm{c}} \ee^{\lambda_{\bm{m},\mathrm{c}} |t|}
       \norm{\bm{\pi}_{\bm{m}}^\mathrm{c}(\bm{x})},\\
      &\forall t\ge 0,\ \bm{x}\in E_{\bm{m}}^\mathrm{s}&:\nonumber\\
      &\norm{D_{\bm{m}}\bm{\phi}^t \bm{\pi}_{\bm{m}}^\mathrm{s}(\bm{x})}
      &\leq c_{\bm{m},\mathrm{s}} \ee^{-\lambda_{\bm{m},\mathrm{s}} t}
       \norm{\bm{\pi}_{\bm{m}}^\mathrm{s}(\bm{x})},\\
      &\forall t\le 0,\ \bm{x}\in E_{\bm{m}}^\mathrm{u}:&\nonumber\\
      &\norm{D_{\bm{m}}\bm{\phi}^t \bm{\pi}_{\bm{m}}^\mathrm{u}(\bm{x})}
      &\leq c_{\bm{m},\mathrm{u}} \ee^{\lambda_{\bm{m},\mathrm{u}} t}
       \norm{\bm{\pi}_{\bm{m}}^\mathrm{u}(\bm{x})}.
    \end{eqnarray}
  \end{enumerate}

  For the sake of simplicity,
  we restrict the constants $\lambda_{\bm{m},x}\ (x=\mathrm{c,s,u})$
  to the absolute condition:\cite{Hirsch1977}
  $\lambda_x=\sup_{\bm{m}}\lambda_{\bm{m},x}$.
    Under this absolute condition,
    the condition 3 for $\bm{\phi}^t\in C^k$ can be rewritten
    with a gap $r\ge 1$ by
    $-\lambda_\mathrm{s}<-r\lambda_\mathrm{c}\le 0
    \le r\lambda_\mathrm{c}<\lambda_\mathrm{u}$.
    Then,
  $\mathcal{M}_\mathrm{NHIM}$ is $C^k$ for some $k\le r$,
  and the manifold is called an {\it eventually and absolutely\/} $k$-NHIM.\cite{Eldering2013}
  For this $k$-NHIM,
  there exists
  a vector field of $\tilde{\bm{\phi}}^t$ and
  a manifold $\tilde{\mathcal{M}}_\mathrm{NHIM}$
  that are $C^k$--close,
  that is, the manifold remains a $k$-\ac{NHIM} or persists under $C^k$ perturbation.

  The theorem was extended to non-compact NHIMs.\cite{Eldering2013}
  In this case, $\mathcal{Q}$ is a bounded geometry
  and $C^k$ is accordingly changed into $C^{k.x}_{b,u}$.
  (See Ref.~\onlinecite{Eldering2013} for those definitions.)

\section{The \ac{NHIM} Persistence Theorem for \acp{RDS}}\label{apx:NHIM_RDS}

  For each $d$-dimensional stochastic path instance $\bm{B}_t$,
  one can uniquely obtain 
  a $d$-dimensional, $t$-continuous function $\bm{\omega}(t)$
  ---as defined in Appendix~\ref{apx:RDEfromSDE})---
  that is associated with the saddle point
  at each instance of time $t$, and for which we are free to 
  initialize at $\bm{\omega}(0)=0$.
  $\bm{\omega}(t)$ is a generalization of the so-called
  \ac{TS} trajectory\cite{dawn05a} discussed in the main text.
  A probability distribution can then be defined for the 
  bundle of instances $\bm{\omega}$ from the probability distribution of $\bm{B}_t$,
  Thus, the event space $\Omega$, which contains this bundle,
  is defined by
  the corresponding $d$-dimensional $t$-continuous functions
  $C(\mathbb{R},\mathbb{R}^d)$:
  \begin{equation}
    \Omega=\left\{\bm{\omega}\middle|
    \bm{\omega}\in C(\mathbb{R},\mathbb{R}^d), \bm{\omega}(0)=0
    \right\}.
  \end{equation}
  We define the
  Wiener shift $\theta_s$,
  and
  a cocycle
  $\bm{\phi}^t_{\bm{\omega}}$
  ---that is a random invariant manifold $\mathcal{M}(\bm{\omega})$---
  such that,
  $\bm{\phi}^t_{\bm{\omega}}(\mathcal{M}(\bm{\omega}))
  =\mathcal{M}(\theta_t\bm{\omega})$.
  The persistence of
  the compact \ac{NHIM} theorem for \ac{RDS} holds\cite{Li2013}
  for these invariant manifolds.
  Let us enumerate the differences in the expression of the persistence theorems
  between Refs.~\onlinecite{Eldering2013,Li2013}.
  \begin{enumerate}
    \item
      The following are all $\bm{\omega}$ dependent:
      $\mathcal{W}_\mathrm{NHIM}^\mathrm{(s)}(\bm{\omega})$,
      $\mathcal{W}_\mathrm{NHIM}^\mathrm{(u)}(\bm{\omega})$,
      $E^x(\bm{\omega})$,
      $\bm{\pi}^x_{\bm{m}}(\bm{\omega})$,
      $c_{\bm{m},x}(\bm{\omega})$,
      and
      $\lambda_{\bm{m},x}(\bm{\omega})$
      $(x=\mathrm{s, u, c})$,
      as well as
      $\mathcal{M}_\mathrm{NHIM}(\bm{\omega})$,
      and $\bm{\phi}^t_{\bm{\omega}}$.
    \item
      Eq.~\eqref{eq:splitting} holds only for
      $E^x~(x=\mathrm{c,u})$.
      For $E^\mathrm{s}$,
      \begin{equation}
        D_{\bm{m}}\bm{\phi}^t_{\bm{\omega}}(E^\mathrm{s}_{\bm{m}}(\bm{\omega}))
        \subset
        E^\mathrm{s}_{D_{\bm{m}}\bm{\phi}^t_{\bm{\omega}}(\bm{m})}(\theta_t\bm{\omega}).
      \end{equation}
    \item (Persistence)
    $\mathcal{M}_\mathrm{NHIM}(\bm{\omega})$ is
    a $1$-normally hyperbolic random invariant manifold.
  \end{enumerate}
  However, $\mathcal{M}_\mathrm{NHIM}(\bm{\omega})$ is still
  $C^k~(r\ge k)$ smooth
  when $\bm{\phi}^t_{\bm{\omega}}$ is $C^k$ and
  $-\lambda_{\mathrm{s}}(\bm{\omega})<-r\lambda_{\mathrm{c}}(\bm{\omega})<
  0<r\lambda_{\mathrm{c}}(\bm{\omega})<\lambda_{\mathrm{u}}(\bm{\omega})$
  for $\lambda_x(\bm{\omega})=\sup_{\bm{m}}\lambda_{\bm{m},x}(\bm{\omega})$.
  However, these differences do not mean that
  the persistence in the NHIM theorem for \ac{RDS} does not hold  
  for more relaxed conditions.

\section{\ac{RDE} from \ac{SDE}}\label{apx:RDEfromSDE}
  For the Langevin type \ac{SDE}s, one can obtain its \acp{RDE} from the use of
  a stationary orbit\cite{Duan2015}.
  Suppose, \iac{SDE} with a Wiener process $W_t\sim\mathcal{N}(0,t)$,
  \begin{equation}
    \dd X_t = (a X_t + b(X_t))\dd t + c \dd W_t,
    \label{eq:LangevinTypeSDE}
  \end{equation}
  is transformed by a stationary orbit $\eta_t$, such that,
  \begin{equation}\label{eq:stationary_orbit}
    \dd \eta_t = a \eta_t\dd t + c \dd W_t,
  \end{equation}
  thus for $t_0\le t\le t_1$,
  \begin{flalign}
    &\dv*{(X_t-\eta_t)}{t} = (a x + b(X_t)),\\
    &\eta_t =
    \begin{cases}{}
      \ee^{a(t-t_0)}\eta_{t_0} + c\int_{t_0}^t \ee^{-a(s-t)}\dd W_s\\
      \ee^{a(t-t_1)}\eta_{t_1} - c\int_t^{t_1} \ee^{-a(s-t)}\dd W_s\\
    \end{cases},\label{eq:SOsolution}
  \end{flalign}
  where we use It{\^{o}}'s lemma:
  $\dd (\ee^{-a t}\eta_t)
  =\dd (\ee^{-a t})\eta_t +\ee^{-at}\dd \eta_t+\dd (\ee^{-at})\dd \eta_t
  =\ee^{-a t}(-a \dd t +\dd \eta_t)$,
  and we only wrote terms with order $\dd t$ or less.
  When $x=X-\eta$
  such that $x(t)=X_t$ and $\omega(t):=\eta_t$ with $\eta_0=0$,
  one obtain a corresponding \ac{RDE},
  \begin{equation}
    \dot{x} = (a x + b(x+\theta_t\omega)).
  \end{equation}
  For this equation, one have the corresponding \ac{RDS}.
  The similar discussion can be made
  for the higher dimensional, Langevin type (Eq.~\eqref{eq:LangevinTypeSDE})  systems.
  Here one can observe the \ac{TS} trajectory $\eta_t^\ddagger$,
  \begin{equation}
    \eta_t^\ddagger=
    \begin{cases}{}
      -c\int_{-\infty}^t \ee^{-a(s-t)}\dd W_s~(a<0)\\
      c\int_t^{\infty} \ee^{-a(s-t)}\dd W_s~(a>0)
    \end{cases},
  \end{equation}
  is a special case of the stationary orbit Eq.~\eqref{eq:stationary_orbit}
  by replacing $t_0\to-\infty, t_1\to\infty$ in Eq.~\eqref{eq:SOsolution}.
\section{Sampling the Reactivity Boundary}\label{apx:Sampling}
  \begin{figure}[t]
    \includegraphics[clip=true,width =\figurewide]{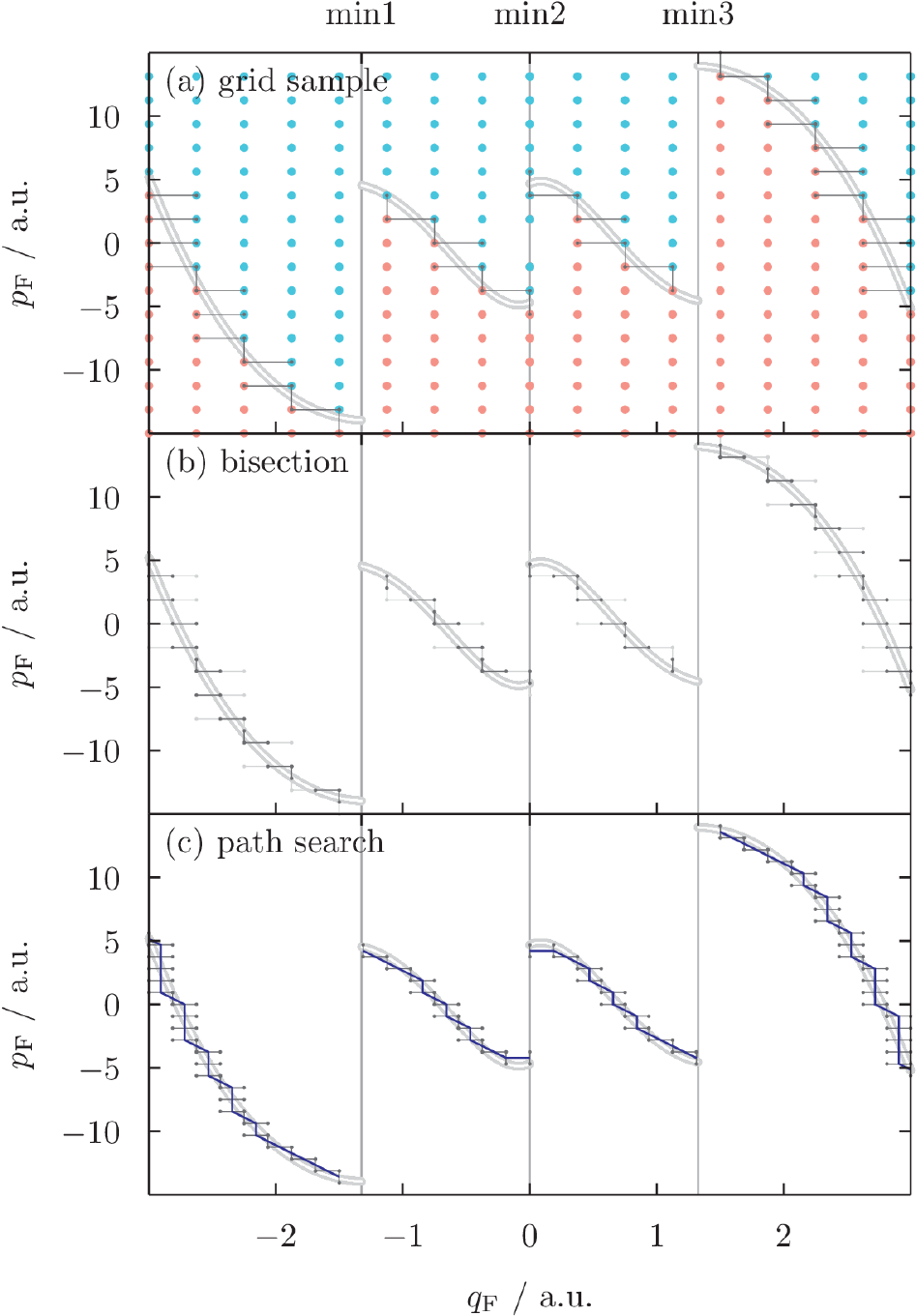}
    \caption{%
      Step-by-step visualization of the algorithm to
      sample neighboring points of the forward-time reactivity boundaries
      in ketene 1 D with external force.
      The solution lines are shown in light gray.
      Straddling pairs of the forward-time reactivity boundaries
      (gray links with dots).
      (a) The initial conditions absorbed at right (left)
      of each cell divided by the coordinate position of
      $q_\mathrm{F}=-3$, min1, min2, min3, and $q_\mathrm{F}=3$
      after positive time integration (blue (red) dots).
      (b) The initial straddling pairs
      before (after) the bisection method applied (light gray (dark gray)).
      (c) The result of the path search algorithm
      (see Fig.~\ref{fig:Algorithm})
      and its midpoint approximation of the neighboring points (blue).
    }\label{fig:search}
  \end{figure}

  In this subsection, we demonstrate how the algorithm,
  introduced in Sec.~\ref{sec:Method} and Fig.~\ref{fig:Algorithm}, works
  in 1 D ketene with external force (Eq.~\eqref{eq:EoM1D}).
  Ideally, it is better to just locate the asymptotic trajectories,
  and this can be achieved by the perturbation theories.
  However, for irregular reactions,
  there could be a case for which the theories are not applicable,
  and need to use numerical investigation
  until an applicable theory is developed.

  The key step in
  the algorithm is the minimization of the uniform sampling.
  In Fig.~\ref{fig:search}, we visualize the locating process.
  Fig.~\ref{fig:search}a,
  shows initial conditions by $(2^4+1)\times(2^4+1)$ grid sampling.
  This produce more than one sample for each area divided by the reactivity boundaries
  thus the sampling is sufficient as seeds for the algorithm.
  In the next step (Fig.~\ref{fig:search}b),
  we apply the bisection method until
  when the demanded number of samples are obtained from the resolution.
  In the figure,
  the bisection method is only applied once for the visualization purpose.
  Then in the bottom figure, the results of the search on the resolution is
  given by the use of the algorithm we introduced in Fig.~\ref{fig:Algorithm}.
  Finally,
  one can apply the bisection method for each pair
  until when the demanded precision is achieved.

  \begin{figure}[t]
    \includegraphics[clip=true,width =\figurewide]{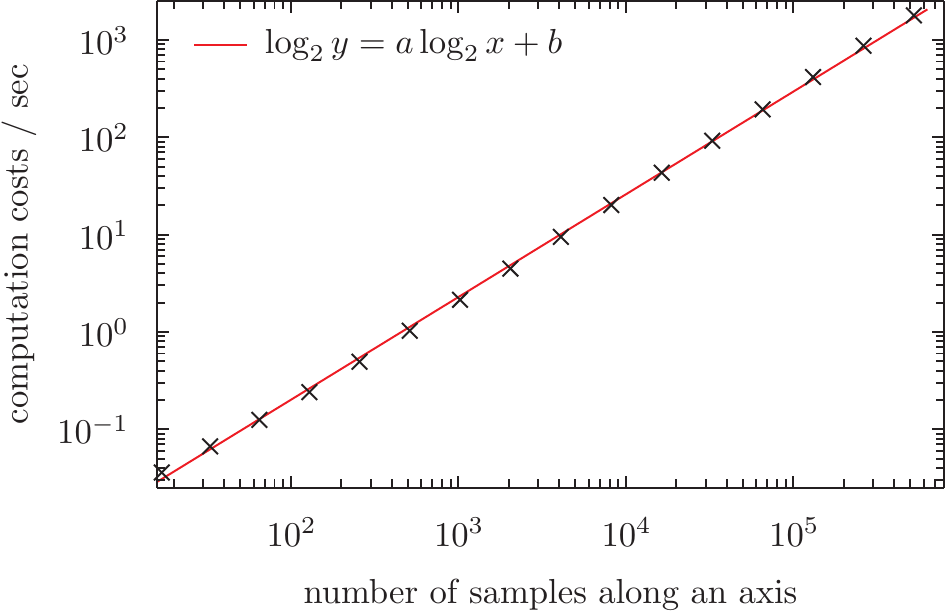}
    \caption{%
      Computation cost in seconds as a function of the sampling resolution
      (see also Fig.~\ref{fig:search}).
      The initial grid size is $(2^4+1)\times (2^4+1)$.
      The first bisection is applied up to $(2^{N_\mathrm{res}}+1)\times (2^{N_\mathrm{res}}+1)$ grid resolution
      where $(2^{N_\mathrm{res}}+1)$ is value of the $x$-axis.
      After the path search, the bisection method is
      applied again up to $(2^{30}+1)\times(2^{30}+1)$ grid resolution.
      The results are shown as cross symbols.
      The least square fit is
      $\log_2{y}=1.0558\times\log_2{x}+9.3353$
      (red line).
    }\label{fig:cost}
  \end{figure}

  Fig.~\ref{fig:cost} shows the computational costs
  of the different sampling resolutions.
  That is,
  when we chose the sampling resolution $(2^{N_\mathrm{res}}+1)\times (2^{N_\mathrm{res}}+1)$
  (Fig.~\ref{fig:search}-bisection),
  and chose the precision achieved by $(2^{30}+1)\times (2^{30}+1)$ grid,
  the $x$--axis of the figure is given by $2^{N_\mathrm{res}}+1$, and $y$--axis is
  given by actual computational time observed by \texttt{std::clock},
  which is implemented in the C++ standard template library.
  In the $\log$-$\log$ plot,
  the cost is an almost linear-order increase over the sampling resolution.
  The order is able to estimate
  by the parameter $a\approx 1$ of the fitting function
  $\log_{2}{y}=a \log_{2}{x}+b$ where $(a,b)=(1.0558,9.3353)$.
  Hence, the algorithm effectively decreases one polynomial order of
  the computational cost from uniform samplings.
  This is more efficient than the \ac{LD} calculation.
  Notice that the computation cost for the worst case is still second order
  because one does not know the shape of the reactivity boundary before hands
  and there is a case that the boundaries are densely existed.

  \begin{figure}[t]
    \includegraphics[clip=true,width =\figurewide]{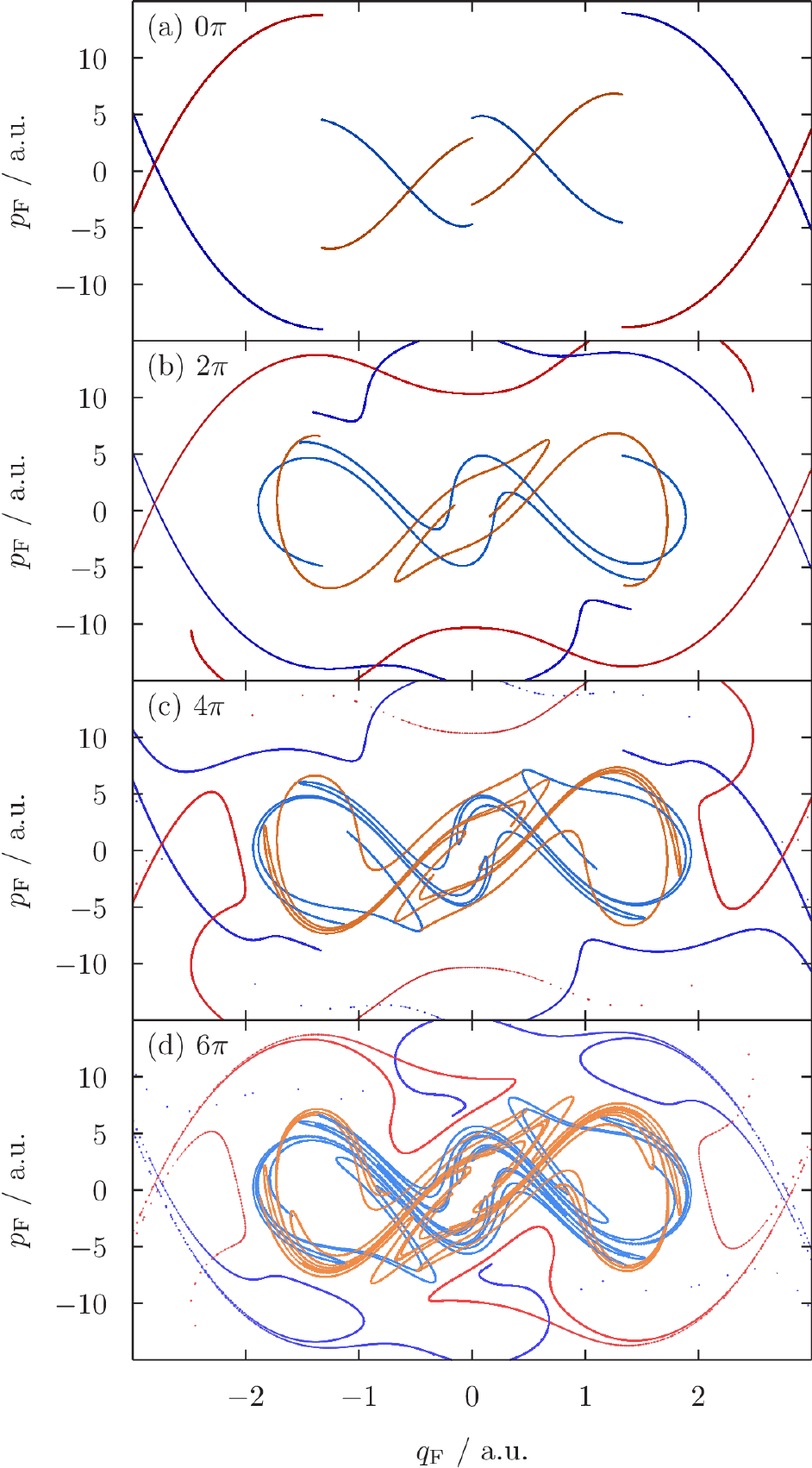}
    \caption{%
      The reactivity boundaries
      at phase $|\omega t|=0, 2\pi, 4\pi, 6\pi$ (a-d)
      obtained by a method explained in Fig.~\ref{fig:search}
      with the resolution given by $N_\mathrm{res}=16$.
      The reactivity boundaries are
      corresponding to stable (blue, cyan) and unstable (red, orange) manifolds.
      The colors of the manifolds get lighter as $|t|$ increases.
    }\label{fig:steps}
  \end{figure}

  If the external field is periodic,
  for the frequency $\omega$ of the periodic field,
  the phase space at $t=0$ is identical when $\omega t \equiv 0 \pmod{2\pi}$.
  To take this advantage, one can propagate time $t$ and get more samples.
  In Fig.~\ref{fig:steps},
  we depict the results of time ($|\omega t|=0, 2\pi, 4\pi, 6\pi$) integrations
  of the straddling pairs of manifolds obtained by the algorithm.
  In the figure,
  the propagated sample points (gray), midpoint estimation of each pair,
  and the line between the estimations
  (red or orange for $\tau_\mathrm{b}$, blue or cyan for $\tau_\mathrm{f}$)
  are shown.
  The strength of each color is proportional to the phase $|\omega t|$.
  For example for the phase $|\omega t|=2\pi M$, $M=0$ is darkest,
  and the color get lighter for larger $M$.
  We employ  $N=12$
  for the sampling resolution $(2^{N_\mathrm{res}}+1)\times (2^{N_\mathrm{res}}+1)$ of the algorithm.
  The gray dots and lines are not visible
  because the estimation is precise,
  and those are overwritten by midpoint estimations.
  The lines between the midpoints are given
  when the two consecutive midpoints have
  less than a certain distance on the figure.
  The disconnection appears from the phase $M\geq 2$
  indicating that the part of the manifold has insufficient samples.
  In the other visualizations,
  we only show the lines for the sake of simplicity.

  \begin{figure}[t]
    \includegraphics[clip=true,width =\figurewide]{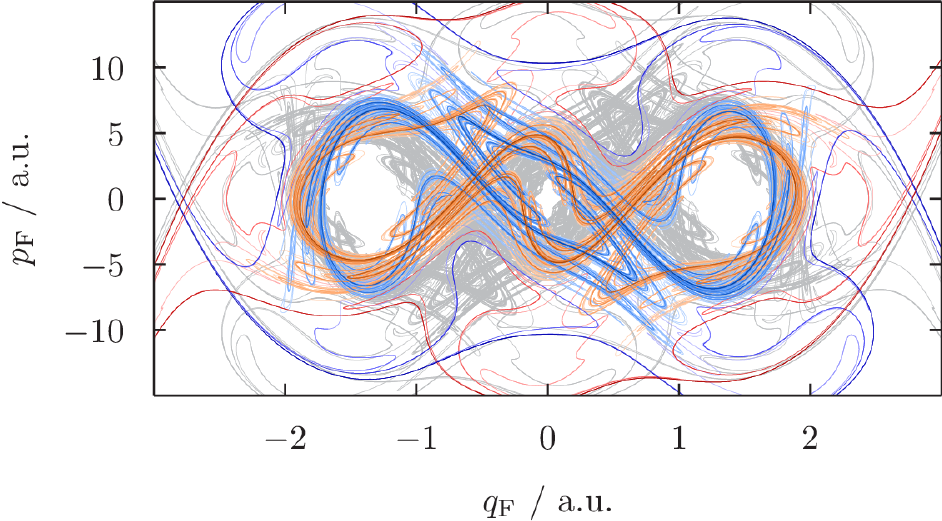}
    \caption{%
      The stable (unstable) manifolds
      at phase $|\omega t| \equiv \pi~\pmod{2\pi}$ (blue, cyan (red, orange)).
      A comparison can be made with another phase
      $|\omega t| \equiv 0~\pmod{2\pi}$ (gray)
      in the background.
    }\label{fig:pi}
  \end{figure}

  The periodical identity also can be used for other phases,
  {\it e.g.}\ for $|\omega t| \equiv \pi \pmod{2\pi}$.
  In Fig.~\ref{fig:pi},
  we show the manifolds at
  the phase $|\omega t| = \pi+2\pi M(M=0,\dots,6)$
  (colored as the same manner as in Fig.~\ref{fig:steps})
  superimposed on the phase $|\omega t|= 2\pi M (M=0,\dots,7)$ (gray).
  Similarly,
  an arbitrary intermediate phase can be obtained.

\newpage
\bibliography{q18bib}
%\bibliography{j,japs,hern,tst,bio,noneq,polymer,liquid,voth,asmd2,papi,gas,osc-bar,roaming,halcyon-X2,miller2,cvpt,yutaka}

\end{document}